\documentclass[11pt]{article}

\usepackage{natbib}

\usepackage{catchfilebetweentags}
\usepackage{geometry}
\usepackage{hyperref}
\usepackage{jcd}
\usepackage[ruled,linesnumbered]{algorithm2e}

\usepackage[mathscr]{euscript}

\newcommand{\err}{\mathtt{err}}

\newcommand{\eps}{\varepsilon}

\renewcommand{\(}{\left(}
\renewcommand{\)}{\right)}

\newcommand{\lbr}{\left\{}
\newcommand{\rbr}{\right\}}

\newcommand{\Ep}{\mathbb{E}}

\renewcommand{\P}{\mathbb{P}}

\newcommand{\absv}[1]{\left|{#1}\right|} 

\newcommand{\1}[1]{\mathbf{1}\lbr #1\rbr}

\newcommand{\spn}[1]{\text{span}\(#1\)}

\newcommand{\cW}{\mathcal{W}}
\newcommand{\cX}{\mathcal{X}}
\newcommand{\cY}{\mathcal{Y}}
\newcommand{\cZ}{\mathcal{Z}}

\newcommand{\est}[1]{\widehat{#1}}

\definecolor{innerboxcolor}{rgb}{.9,.95,1}
\definecolor{outerlinecolor}{rgb}{.6,0,.2}
\definecolor{shcolor}{RGB}{27, 87, 14}
\definecolor{rckcolor}{RGB}{0,0,255}

\newcommand{\txt}[1]{\textup{#1}}
\newcommand{\iid}{\overset{\text{i.i.d.}}{\sim}}

\newcommand{\ti}[1]{\tilde{#1}}

\newcommand{\dpsd}{d_{\textup{psd}}}
\newcommand{\colspace}{\mathsf{Col}}
\newcommand{\genpsd}{A}

\renewcommand{\S}{\mathbb{S}}

\newcommand{\Hone}{\mathtt{H}_1^n}
\newcommand{\Htwo}{\mathtt{H}_2^n}
\newcommand{\Hell}{\mathtt{H}_{\ell}^n}

\newcommand{\dham}{d_{\txt{ham}}}

\newcommand{\cV}{\mathcal{V}}

\newcommand{\alg}{\hyperref[alg:fast]{\mathtt{ALG}}}
\newcommand{\computeS}{\hyperref[alg:compute-s]{\mathtt{ComputeS}}}
\newcommand{\preprocess}{\hyperref[alg:preprocess]{\mathtt{Preprocess}}}

\newcommand{\SigmaAlg}{\Sigma_{\mathtt{ALG}}}

\newcommand{\bnorm}{b_{\mathsf{norm}}}
\newcommand{\binner}{b_{\mathsf{inner}}}
\newcommand{\mop}{m_{\mathsf{op}}}
\newcommand{\aop}{\alpha_{\mathsf{op}}}
\newcommand{\tnorm}{t_{\mathsf{norm}}}
\newcommand{\tinner}{t_{\mathsf{inner}}}
\newcommand{\mip}{m_{\mathsf{res}}}
\newcommand{\mbig}{m_{\mathsf{inner}}}
\newcommand{\tclip}{t_{\mathsf{clip}}}

\newcommand{\ds}{\cY}
\newcommand{\uds}{\cZ}
\newcommand{\cds}{\cX}

\newcommand{\scov}[2]{\Sigma(#1_{#2})}
\newcommand{\scovinv}[2]{\Sigma(#1_{#2})^{\dag}}
\newcommand{\scovisqrt}[2]{\Sigma(#1_{#2})^{\dag/2}}
\newcommand{\scovsqrt}[2]{\Sigma(#1_{#2})^{1/2}}
\newcommand{\wmax}{w_\txt{max}}
\newcommand{\Sgone}{Q}
\newcommand{\Sgtwo}{Q'}

\newcommand{\genset}{H}
\newcommand{\ipresset}{F}

\newenvironment{proof-of-corollary}[1][{}]{\noindent{\bf Proof of Corollary {#1}}
  \hspace*{1em}}{\qed\bigskip\\}

\newcommand*{\SectionsPath}{sections/}

\begin{document}

\begin{center}
    \Large{A fast and slightly robust covariance estimator} \\
    \vspace{.5cm}

    \large{John Duchi$^{1,2}$ ~~~~ Saminul Haque$^3$ ~~~~ Rohith Kuditipudi$^3$} \\
    \vspace{.25cm}
    \large{Departments of $^1$Statistics, $^2$Electrical Engineering,
        and $^3$Computer Science \\
        Stanford University}
    \\
    \vspace{.2cm}
    \large{February 2025}
\end{center}

\begin{abstract}
    Let $\cZ = \{Z_1, \dots, Z_n\} \simiid P \subset \R^d$ from a distribution $P$
    with mean zero and covariance $\Sigma$.
    Given a dataset $\cX$ such that $\dham(\cX, \cZ) \leq \eps n$, we are interested
    in finding an efficient estimator $\widehat{\Sigma}$ that achieves
    $\err(\widehat{\Sigma}, \Sigma)
        \defeq
        \|\Sigma^{-\frac{1}{2}}\widehat{\Sigma}\Sigma^{-\frac{1}{2}} - I\| \leq 1/2$.
    We focus on the low contamination regime $\eps = o(1/\sqrt{d}$).
    In this regime, prior work required either $\Omega(d^{3/2})$ samples or runtime that is
    exponential in $d$.
    We present an algorithm that, for subgaussian data, has near-linear
    sample complexity $n = \wt{\Omega}(d)$ and runtime
    $O((n+d)^{\omega + \frac{1}{2}})$, where $\omega$ is the matrix
    multiplication exponent.
    We also show that this algorithm works for heavy-tailed data with near-linear
    sample complexity, but in a smaller regime of $\eps$.
    Concurrent to our work, \cite{DiakonikolasHoPeTi24b} give Sum-of-Squares estimators
    that achieve similar sample complexity but with large polynomial runtime.
\end{abstract}

\section{Introduction}
\subsection{Problem formulation}\label{sec:problem-formulation}
Let $\cZ = \{Z_1,...,Z_n\} \iid P$ for a distribution $P$ over $\R^d$
with mean zero
and covariance $\Sigma = \Ep_P[(Z-\mu)(Z-\mu)^T]$,
and let the observations $\cX \in (\R^d)^n$ satisfy $\dham(\cX,\cZ) \leq \eps n$
almost surely, i.e., $\cX$ is an \textit{$\eps$-contamination} of $\cZ$.
Defining the relative operator error
\begin{align*}
  \dpsd(A, B) \defeq
  \begin{cases}
    \norm{A^{\dag/2} (B - A) A^{\dag/2}}_\txt{op}
           & \mbox{if}~\colspace(B) \subseteq  \colspace(A) \\
    \infty & \text{otherwise},
  \end{cases}
\end{align*}
we design an efficient algorithm $\alg$ to compute an estimate $\est{\Sigma} = \alg(\cX)$ that
minimizes
\begin{align}
  \err(\est{\Sigma}, \Sigma) \defeq \dpsd(\est{\Sigma},\Sigma). \label{eqn:error-defn}
\end{align}
We apply this algorithm to the following two families of distributions.

\begin{definition}[Hypercontractive subgaussian]\label{def:hyper-subg}
  A distribution $P$ over $\R^d$ with mean zero and covariance
  $\Sigma$ is $\sigma^2$-hypercontractive subgaussian if
  for all $v \in \R^d$
  \begin{align*}
    \E_{Z \sim P}[\exp(Z \cdot v)] \leq e^{\sigma^2 v^T \Sigma v/2}.
  \end{align*}
\end{definition}

\begin{theorem}[Informal]\label{thm:subg-result-informal}
  Let $n = \wt{\Omega}(d)$ and $\eps = \wt{O}(d^{-1/2})$.
  Let $P$ satisfy Definition~\ref{def:hyper-subg} with $\sigma = O(1)$.
  Then $\err(\alg(\cX), \Sigma) \leq 1/2$ with probability at least $3/4$.
\end{theorem}

\begin{definition}[Moment bounded]\label{def:moment-bounded}
  A distribution $P$ over $\R^d$ with mean zero and covariance
  $\Sigma$ is $k$-th moment bounded by $m_k$ if for all $v \in \R^d$
  \begin{align*}
    \E_{Z \sim P}[|Z \cdot v|^k]^{1/k} \leq m_k (v^T \Sigma v)^{1/2}.
  \end{align*}
\end{definition}

\begin{theorem}[Informal]\label{thm:heavy-tail-result-informal}
  Let $n = \wt{\Omega}(d)$ and $\eps = \wt{O}\left(d^{-(\frac{1}{2}+\frac{1}{k})}\right)$.
  Let $P$ satisfy Definition~\ref{def:moment-bounded} with $k \geq 4$ and $m_k = O(1)$.
  Then $\err(\alg(\cX), \Sigma) \leq 1/2$ with probability at least $3/4$.
\end{theorem}

Prior to our work, the best known polynomial time estimators required $n = \Omega(\eps d^2)$
even for subgaussian data \citep{DiakonikolasKa22,KothariStSt18}.
Yet, information-theoretically only $d$ samples are necessary so long as $\eps$ is a sufficiently small constant.
For contamination rates $\eps \ll d^{-1/2}$ (depending on the degree to which $P$ concentrates), our algorithm achieves this optimal sample complexity
up to polylogarithmic factors in $d$.
Its runtime is $(n+d)^{\omega + 1/2}$ assuming $\omega$ is the matrix multiplication exponent.
Thus, we substantially improve upon the sample complexity of existing polynomial time algorithms
in the small contamination regime where $d^{-1} < \eps \ll d^{-1/2}$.

In recent work concurrent to ours, \citet{DiakonikolasHoPeTi24b} construct a family of estimators based on the Sum-of-Squares (SoS) hierarchy
that achieve sample complexity $n = \wt{\Omega}(\eps^2 d^{1+\gamma})$ and runtime $(n+d)^{O(1/\gamma)}$ assuming
$P$ is hypercontractive subgaussian. In the regime where $\eps \ll d^{-1/2}$, their results imply a polynomial time SoS estimator
with sample complexity $n = \wt{O}(d)$ like ours. However, the runtime of this estimator is substantially worse than what we achieve.
We defer a full discussion of related work to Section~\ref{sec:related-work}.

We describe our estimator in Algorithm~\ref{alg:fast} ($\alg$) and give a high-level overview
of its design in Section~\ref{sec:algorithm}. In Section~\ref{sec:goodness-def}, we define properties of a dataset $\cZ$
that collectively determine the degree to which $\cZ$ is a \textit{good} dataset (Definition~\ref{defn:good-dataset}).
In Section~\ref{sec:goodness-def}, we prove $\alg$ robustly estimates of $\cZ$
from observations $\cX$ that are an $\eps$-contamination of $\cZ$ so long as $\cZ$ is good.
Then, we give bounds for $n$ and $\eps$ under which i.i.d.\ samples from subgaussian and moment bounded distributions
are good with high probability in Sections~\ref{sec:subg-perf} and~\ref{sec:heavy-perf}, respectively.
These bounds establish the robustness and sample complexity of $\alg$ for these
classes of distributions, thus formalizing Theorems~\ref{thm:subg-result-informal} and \ref{thm:heavy-tail-result-informal}.

\subsection{Related work}\label{sec:related-work}
Despite the study of robust statistical estimation dating back to
more than a half-century ago \citep{Tukey60,Huber64}, only in the last decade have researchers developed
computationally efficient algorithms with optimal or near-optimal sample complexity.
Originating from the work of \citet{DiakonikolasKaKaLiMoSt16} and \citet{LaiRaVe16},
we now have polynomial time algorithms for robustly estimating the mean
of an isotropic distribution to within small $\ell_2$ error from
$n = \wt{\Omega}(d)$ observations and robustly estimating the covariance of a fourth moment bounded distribution
to within small relative Frobenius error from $n = \wt{\Omega}(d^2)$ observations, even when a constant fraction of these
observations are contaminated.
In many cases, the runtimes of these algorithms are not just polynomial but in fact quite fast \citep{ChengDiGeWo19,ChengDiGeSo20,DongHoLi19,DepersinLe19,LiYe20} (see the book of \citet{DiakonikolasKa22} for a more thorough survey).

For anisotropic distributions, though
it is information theoretically possible to robustly estimate the covariance of a moment bounded distribution to within small
relative operator error~\eqref{eqn:error-defn} from $n = \Omega(d)$ observations \citep{MendelsonZh18,LugosiMe20},
prior to our work
all known polynomial time estimators required $n = \Omega(\eps d^2)$ observations.
Thus, even for $\eps = \omega(d^{-1})$, robustly estimating of the mean of an bounded covariance distibution
to within small covariance-normalized error required either $\exp(d)$ runtime or $n = \omega(d)$ observations.
For $\eps \gg d^{-1/2}$ and subgaussian data, the existence of a polynomial time sample efficient estimator remains an open question;
for moment bounded distributions, the existence of a polynomial time estimator with sub-quadratic sample complexity $n = o(d^2)$
remains open.

Independently and concurrently to our work, \citet{DiakonikolasHoPeTi24b} derive SoS certificates of
operator norm resilience: i.e., SoS proofs of the stability of a dataset's covariance
in operator norm with respect to removing any $\eps$ fraction of the data.
They are then are able to convert these certicates into robust covariance estimators using standard techniques
(see e.g., \citet{KothariStSt18}).
Their estimators make trade-offs between $n$, $\eps$ and runtime; in particular, on subgaussian data
for $n = \wt{\Omega}(\eps^2 d^2)$ they achieve small relative operator error in quasipolynomial time.
They also mention a folklore polynomial time SoS certificate with sample complexity $n = \wt{\Omega}(d)$ for $\eps \ll d^{-1/2}$.
In comparison, our estimator fails catastrophically for $\eps = \Omega(d^{-1/2})$ but otherwise achieves the optimal sample complexity with much faster runtime.
Our estimator also works on moment bounded distributions (albeit for smaller $\eps$).

\subsection{Notation}\label{sec:notation}

\paragraph{Semidefinite matrices and norms}

\newcommand{\Proj}{\Pi}  

For a positive semidefinite (PSD) matrix $\genpsd \in \R^{d\times d}$, we
let $\colspace(\genpsd)$ denote its columnspace,
$\genpsd^\dag$ its pseudoinverse,
and $\genpsd^{\dag/2}$ the square-root of the pseudoinverse.

Define the extended-value inner product corresponding to $\genpsd \succeq 0$ between $u,v \in \R^d$ by
\begin{equation*}
  \langle u, v\rangle_\genpsd \defeq
  \lim_{t \downarrow 0}
  v^T(\genpsd + t I)^{-1} u
  =
  \begin{cases}
    v^T \genpsd^\dag u & u,v \in \colspace(\genpsd) \\
    +\infty            & \text{otherwise}
  \end{cases}
\end{equation*}
and the extended-value Mahalanobis
norm $\norm{\cdot}_\genpsd$ by $\norm{v}_\genpsd^2 = \langle v, v\rangle_\genpsd$.
This norm has the monotonicity property that
if $\genpsd \preceq B$, then $\norm{v}_\genpsd \geq \norm{v}_{B}$
for all $v \in \R^d$.
When $\genpsd$ is non-singular, we have
$\norm{v}_\genpsd^2 = v^T A^{-1} v$ and $\langle u, v\rangle_\genpsd = u^T A^{-1} v$.
We abuse notation throughout the paper to equate $\norm{v}_\genpsd^2 = v^T A^\dag v$ and $\langle u, v\rangle_\genpsd = u^T A^\dag v$
in general.
Also, for $A,B \succeq 0$ we use $A^\dag \preceq B^\dag$ to denote $\norm{v}_A \leq \norm{v}_B$ for all $v \in \R^d$.
Otherwise (e.g., in the context of a generic matrix multiplication operation),
we use $A^\dag$ to denote the usual pseudoinverse.

We use $\norm{\cdot}$ to denote both the $\ell_2$ norm of a vector and
the operator norm of a matrix.

\paragraph{Sets and Covariances}

For a dataset $\cZ = \{z_1, \dots, z_n\} \subseteq \R^d$, let
$\scov{\uds}{} \defeq \frac{1}{n} \sum_{i=1}^n z_i z_i^T$.
For $S \subseteq [n]$, let $\cZ_S$ denote the sub-dataset
$\{z_i\1{i \in S} \mid i \in [n]\}$.
Notice that $\cZ_S$ is still a dataset of size $n$ rather than size $|S|$,
so
$\scov{\ds}{S} = \frac{1}{n}\sum_{i \in S} z_i z_i^T$.
Finally, we use $\Hone = \{1, \dots, n/2\}$ and $\Htwo = \{n/2+1, \dots, n\}$ to denote
the two halves of $[n]$.

\section{Algorithm description}\label{sec:algorithm}

\begin{algorithm}[h]
  \DontPrintSemicolon
  \caption{Robust covariance estimation ($\alg$)}
  \label{alg:fast}
  \setcounter{AlgoLine}{0}
  \KwIn{Dataset $\cX = \{x_1, \dots, x_n\} \subset \R^d$,
    contamination fraction $\eps$,
    norm threshold $\tnorm$,
    inner product threshold $\tinner$}
  \KwOut{Estimated covariance matrix $\Sigma_{\alg}$}
  $R \leftarrow [n]$ \label{alg:R-init} \;

  \While{true \label{alg:main-loop-start}}{
    $R_1 \leftarrow R \cap \Hone$\; \label{alg:R1-init}

    $R_2 \leftarrow R \cap \Htwo$\; \label{alg:R2-init}

    \While{$\exists \ell, \ell' \in \{1, 2\}, i \in R_\ell
        \mathrm{~s.t.~}
        x_i^T \scovinv{\cds}{R_{\ell'}} x_i > \tnorm^2
      $
      \label{alg:norm-filter-start}}
    {
      Remove $i$ from $R_\ell$\; \label{alg:norm-filter-pt-removal}
    } \label{alg:norm-filter-end}

    $S_1 \leftarrow \computeS(\cX_{R_1}, \scov{\cds}{R_2}, \tinner)$\;
    \label{alg:compute-S-call-1}

    $S_2 \leftarrow \computeS(\cX_{R_2}, \scov{\cds}{R_1}, \tinner)$\;
    \label{alg:compute-S-call-2}

    $R \leftarrow R_1 \cup R_2$\; \label{alg:R-def}

    $S \leftarrow S_1 \cup S_2$\; \label{alg:S-def}

    \If{$\norm{\scovisqrt{\cds}{R}\scov{\cds}{S}\scovisqrt{\cds}{R}}
        \leq \frac{\eps \tinner}{4e \log(\tnorm)}$
      \label{alg:S-norm-check}}{
      \Return{$\scov{\cds}{R}$}\; \label{alg:return}
    }
    $v_S \leftarrow
      \argmax_{v \in \S^{d-1}} v^T\scovisqrt{\cds}{R}\scov{\cds}{S}\scovisqrt{\cds}{R}v$\;
    \label{alg:v-def}

    \For{$i \in [n]$}{
      $w_i \leftarrow (x_i^T \scovisqrt{\cds}{R} v_S)^2\1{i \in S}$\; \label{alg:w-def}
    }

    $\wmax \leftarrow \max_{i \in [n]} w_i$\; \label{alg:w-alg-def}

    $\xi \sim \txt{Unif}([0,1])$\; \label{alg:xi-def}

    Remove $i$ from $R$ if $\frac{w_i}{\wmax} \geq \xi$\;
    \label{alg:random-filter}
  }\label{alg:main-loop-end}
\end{algorithm}

\begin{algorithm}[h]
  \DontPrintSemicolon
  \caption{Find large inner product pairs ($\computeS$)}
  \label{alg:compute-s}
  \setcounter{AlgoLine}{0}
  \KwIn{Dataset $y_1, \dots, y_n \in \R^d$,
    PSD matrix $A$,
    inner product threshold $\tinner$}
  \KwOut{Set $S$ of points with large inner products}
  $S \leftarrow \emptyset$\; \label{compS:S-init}

  \While{$\exists i,j \not\in S\ \mathrm{such \ that}\ |y_i^T A^\dag y_j| > \tinner$
    \label{compS:loop-cond}}{
    Add $i$ and $j$ to $S$\; \label{compS:add-to-S}
  }\label{compS:loop-end}
  \Return{$S$ \label{compS:return-S}}
\end{algorithm}

We describe our estimator $\alg$ in Algorithm~\ref{alg:fast} and presently give
an overview of its design.
Recall the input $\cX$ to $\alg$ is an $\eps$-contamination of $\cZ = \{z_1, \dots, z_n\} \iid P$.
Let $G,B \subseteq [n]$ denote the ``good'' and ``bad''
subsets of $\cX$, i.e., $G = \{i \in [n] \mid x_i = z_i\}$ and $B = [n] \setminus G$.
We can decompose $\scov{\cds}{}$ into its good and bad components
so that $\scov{\cds}{} = \scov{\cds}{G} + \scov{\cds}{B}$.
The goal of $\alg$ will be to remove elements from $R \subseteq [n]$ until
$\scov{\cds}{B \cap R}$ is small, at which point (so long as we do not remove too many good data)
the remaining covariance $\scov{\cds}{R \cap G}$ will be close to $\scov{\cds}{G}$ and therefore
a good estimate of the uncontaminated covariance $\scov{\uds}{}$.

Suppose $\scov{\cds}{B \cap R}$ is not small relative to $\scov{\cds}{R}$ in the current iteration of $\alg$, i.e.,
there exists a ``bad direction'' $v \in \S^{d-1}$ such that $v^T \scovisqrt{\cds}{R} \scov{\cds}{B \cap R} \scovisqrt{\cds}{R} v = \Omega(1)$,
and let $B_v = \{i \in B \cap R \mid v^T \scovisqrt{\cds}{R} x_i \geq C/\sqrt{\eps}\}$.
Then for sufficiently small constant $C$ it will be the case that $v^T \scovisqrt{\cds}{R} \scov{\cds}{B_v \cap R} \scovisqrt{\cds}{R} v = \Omega(1)$.
This implies so long as we can reliably identify elements in $B_v$ and include these elements in $S$, then
$\alg$ will not terminate while there remain bad directions.
The main idea of $\alg$ is that we can identify elements in $B_v$ based on the fact that the corresponding points should have large inner products with
each other.
In particular, in each iteration of $\alg$ we will add elements to $S$ whose covariance normalized inner products
with at least one other remaining element in $R$ is $\Omega(1/\eps)$.

There are two technical challenges that arise in this process.
The first challenge arises in proving points in $B_v$ have high inner product with each other;
in particular, the contribution to the inner product from the direction $v$ can be masked by the remaining
$d-1$ orthogonal directions. We show via a simple geometric argument
(Lemma~\ref{lemma:negative-correlation-bound})
that enough contaminated points will have
$\Omega(1/\eps)$ covariance normalized inner product with some other
contaminated point to ensure
that $v^T \scov{\cds}{B \cap S} v = \Omega(v^T \scov{\cds}{R} v)$.

The second challenge is to show that $v^T \scov{\cds}{G \cap S} v$ is small.
In particular, while the inner products among uncontaminated data with respect to the uncontaminated sample covariance
should be $o(1/\eps)$ for $\eps \ll 1/\sqrt{d}$,
the contamination can cause the inner product of many uncontaminated points
to exceed $1/\eps$, in which case $\alg$ will include these points in $S$.
We are able to bound (Lemma~\ref{lemma:few-good-points-in-S}) the number of uncontaminated points included in $S$
by $\ti{O}(\eps + \eps^2 d)n$,
which suffices to ensure $v^T \scov{\cds}{G \cap S} v \leq v^T \scov{\cds}{B \cap S} v$
since our assumptions on $P$ imply that any
$o(n)$-size subset of $G$ will have $o(1)$ relative operator norm with respect to $\scov{\cds}{R}$.
This means we will filter more contaminated points that uncontaminated points
(in expectation) until $\alg$ terminates, at which point there will be no directions in
which the contaminated data skews the covariance by too much.

\section{Properties of a good dataset}
\label{sec:goodness-def}
We presently introduce the properties we need the
uncontaminated dataset to satisfy for $\alg$ to work well.
The first property we require is that all points have bounded Mahalanobis norm with respect
to the covariance over the subset $H$ of the uncontaminated dataset, so that the uncontaminated data will not
be filtered in Line~\ref{alg:norm-filter-start} of $\alg$.
\begin{definition}[Bounded norm]\label{def:bdd-norm}
  A dataset $\cZ = \{z_1, \dots, z_n\}$ is $\bnorm$-norm bounded
  if for all $i \in [n]$ and $\genset \in \{\Hone,\Htwo\}$ we have
  $z_i^T \scovinv{\uds}{\genset} z_i \leq \bnorm^2$.
\end{definition}

The second property we require is that most inner products between uncontaminated points be
bounded with respect to uncontaminated covariance over $H$, in order to bound the number of uncontaminated elements
that $\alg$ includes in $S$ (Line~\ref{alg:S-def}).
\begin{definition}[Bounded inner products]\label{def:bdd-ips}
  A dataset $\cZ = \{z_1, \dots, z_n\}$ is $(\mbig, \binner)$-inner product
  bounded if for $\genset \in \{\Hone,\Htwo\}$
  there exists $T \subseteq [n] \setminus \genset$ satisfying $|T| \geq n - |\genset| - \mbig$
  and $|z_i^T \scovinv{\uds}{\genset} z_j| \leq \binner$ for all distinct $i,j \in T$.
\end{definition}

The final property we require is that all sufficiently small
subsets of the data have a covariance that is
bounded above proportionally to the sample covariance.
This will ensure that no small subset has a large effect on the sample
covariance in relative operator norm.
\begin{definition}[Operator norm resilience]\label{def:op-norm-res}
  A dataset $\cZ = \{z_1, \dots, z_n\}$ is $(\mop, \aop)$-operator norm
  resilient if for $\genset \in \{\Hone,\Htwo\}$ we have
  $\scov{\uds}{\genset} \preceq \frac{3}{4} \scov{\uds}{}$ and for all
  $T \subseteq \genset$ with $|T| \leq \mop$ we have
  $\scov{\uds}{T} \preceq \aop \scov{\uds}{\genset}$.
\end{definition}
\begin{lemma}\label{lem:sub-contam-bounded-cov}
  Suppose $\cZ = \{z_1, \dots, z_n\}$ is $(\mop, \aop)$-operator norm
  resilient. Let $\genset \in \{\Hone,\Htwo\}$.
  Then for all $T \subseteq \genset$ such that $|T| \geq |\genset|-\mop$, we have
  \begin{align*}
    (1-\aop)\scov{\uds}{\genset}
    \preceq \scov{\uds}{T}
    \preceq \scov{\uds}{\genset}
    ~~~\text{and}~~~
    \scovinv{\uds}{\genset}
    \preceq \scovinv{\uds}{T}
    \preceq (1-\aop)^{-1} \scovinv{\uds}{\genset}.
  \end{align*}
\end{lemma}
\begin{proof}
  This follows immediately from the definition of operator norm resilience.
\end{proof}

With these properties in hand, we define the notion of a good dataset.
\begin{definition}[Good dataset]\label{defn:good-dataset}
  A dataset $\cZ$ is $(\bnorm, \mbig, \binner, \mop, \aop)$-good
  if it is
  \begin{enumerate}
    \item $\bnorm$-norm bounded;
    \item $(\mbig, \binner)$-inner product bounded; and
    \item $(\mop, \aop)$-operator norm resilient.
  \end{enumerate}
\end{definition}
\section{Analysis of \texorpdfstring{$\alg$}{ALG} for contaminated good data}\label{sec:alg-analysis}

\begin{theorem}\label{thm:alg-result}
  Let $C > 0$ and $\eps, n, \bnorm, \mbig, \binner, \mop, \aop, \tinner, \tnorm$ satisfy
  \begin{enumerate}
    \item $\eps \leq \frac{1}{4(1+C)}$,
    \item $\tnorm \geq (1-\aop)^{-\frac{1}{2}}\bnorm$,
    \item $\tinner \geq \max\{2\binner, \frac{8e\aop\log(\tnorm)}{1-\aop} \cdot \frac{1}{\eps}\}$,
    \item $n \geq \frac{12e\tnorm^2\log(\tnorm)}{\eps\tinner}$,
    \item $\mop \geq \max\{1 + C, 1 + \frac{72(1+C)\tnorm^2}{\tinner}\}\eps n + 4\mbig$.
  \end{enumerate}
  Let $\cZ$ be a $(\bnorm, \mbig, \binner, \mop, \aop)$-good dataset and let
  $\cX$ be an $\eps$-contamination of $\cZ$.
  Then $\alg$ terminates in $T$ iterations with $\Ep[T] \leq 2\eps n$ and returns
  $\Sigma_{\alg}$ satisfying
  \begin{align*}
    (1-\aop)\scov{\uds}{}
    \preceq \SigmaAlg
    \preceq \frac{1+\aop}{1-8\eps\tinner} \scov{\uds}{}
  \end{align*}
  with probability at least $1-C^{-1}$.
\end{theorem}

We presently prove Theorem~\ref{thm:alg-result}.
The bulk of the proof analyzes a single iteration of the outer while-loop
(Section~\ref{sec:single-iteration-analysis});
in particular, we will show that every iteration this loop will either terminate
with the desired utility or will remove more contaminated than uncontaminated elements in expectation.
We will then prove the main result from the single-iteration
result by applying a standard martingale argument (see, e.g., \citet[Theorem 2.17]{DiakonikolasKa22}) to bound the number of removed good elements over all iterations (Section~\ref{sec:thm-pf}).

\subsection{Analysis of one iteration}\label{sec:single-iteration-analysis}

\newcommand{\Rstart}{R^{\mathsf{start}}}
\newcommand{\Rend}{R^{\mathsf{end}}}
\newcommand{\Rnorm}{R^{\mathsf{norm}}}
\newcommand{\Anorm}{\scovisqrt{\cds}{\Rnorm}}

In this section, we analyze a single iteration
of the outer while loop of $\alg$
(Lines~\ref{alg:main-loop-start}-\ref{alg:main-loop-end}).
We show in Proposition~\ref{prop:single-iteration-result} that
assuming not too many uncontaminated points have already
been removed in previous iterations, the algorithm will either terminate and
output an estimate of the sample covariance with the desired error or will remove on average more
contaminated points than uncontaminated points.

Before proceeding, we introduce some notation to track the state of $\alg$ over an iteration.
Let $\Rstart$ be the state of $R$ at the start of the iteration
of the outer while loop
(i.e., before Line~\ref{alg:R1-init} of $\alg$ executes)
and let $\Rend$ be the state of $R$ at the end of the iteration
(i.e., after Line~\ref{alg:random-filter} of $\alg$ executes).
Likewise, let $\Rnorm$ be the state
of $R$ after the norm filtering
(Lines~\ref{alg:norm-filter-start}-\ref{alg:norm-filter-end} of $\alg$)
completes execution.
For all of these states of $R$, we subscript by $\ell \in \{1, 2\}$ to denote
the half of $R$ corresponding to $\Hell$;
e.g., $\Rstart_1 = \Rstart \cap \Hone$.
Finally, we use $G,B \subseteq n$ to track the indices of uncontaminated and
contaminated data, i.e., $G = \{i \in [n] \mid x_i = z_i\}$ and $B = [n] \setminus G$.


\begin{proposition}\label{prop:single-iteration-result}
  Let $\cZ$ and $\cX$ satisfy the conditions of Theorem~\ref{thm:alg-result}.
  Suppose $|G \setminus \Rstart| \leq C\eps n$. Then almost surely either
  \begin{enumerate}
    \item $\alg$ terminates with output $\SigmaAlg$
          satisfying
          $(1-\aop) \scov{\uds}{}
            \preceq \SigmaAlg
            \preceq \frac{1+\aop}{1-8\eps\tinner} \scov{\uds}{}$; or
    \item $\E\left[\left|B \cap (\Rstart \setminus \Rend)\right| \mid \Rstart \right]
            > \E\left[\left|G \cap (\Rstart \setminus \Rend)\right| \mid \Rstart \right]$.
  \end{enumerate}
\end{proposition}

We defer the proof of Proposition~\ref{prop:single-iteration-result} to Appendix~\ref{sec:pf-prop-single-iteration}.
In the remainder of this section, we will state the key helper lemmas we require
and give a sketch of the proof.
Every iteration of $\alg$ begins with a norm-filtering step
(Lines~\ref{alg:norm-filter-start}-\ref{alg:norm-filter-end}).
The first key lemma analyzes the state of $\alg$ after
executing this norm-filtering step.
\begin{lemma}\label{lem:norm-filter-props}
  Suppose $|G \setminus \Rstart| \leq \mop$.
  Then
  \begin{enumerate}
    \item $G \cap \Rstart = G \cap \Rnorm$;
    \item The norm filter terminates in at most $|B \cap \Rstart|+1$ iterations; and
    \item For $i \in \Rnorm$ and $\ell \in \{1,2\}$, we have
          $x_i^T \scovinv{\cds}{\Rnorm_\ell} x_i \leq \tnorm^2$.
  \end{enumerate}
\end{lemma}
\begin{proof}
  Proceeding with the first claim, for $\ell \in \{1,2\}$ we have
  \begin{align*}
    |G \cap \Rstart_\ell| = |\Hell| - |\Hell \cap (G \setminus \Rstart)|
    \geq |\Hell| - \mop.
  \end{align*}
  Thus, for all $i \in G \cap \Rstart$
  \begin{align*}
    x_i^T \scovinv{\cds}{\Rstart_\ell} x_i
     & \leq x_i^T \scovinv{\cds}{G \cap \Rstart_\ell} x_i
    \leq (1-\aop)^{-1} x_i^T \scovinv{\uds}{H_\ell} x_i
    \leq (1-\aop)^{-1} \bnorm^2,
  \end{align*}
  where the second inequality follows from Lemma~\ref{lem:sub-contam-bounded-cov}.
  It follows immediately from the assumption
  $\tnorm \geq (1-\aop)^{-\frac{1}{2}}\bnorm$
  that no uncontaminated elements are filtered the first iteration
  of the norm-filtering step.
  Repeating this argument for all the other iterations, we conclude
  that no uncontaminated elements are filtered throughout.
  This implies the first claim of the lemma, provided the norm-filtering step terminates.

  We now bound the number of iterations before the norm-filtering step terminates.
  Note that for all but the final
  iteration of the norm-filtering step, $\alg$ removes at least one element. Our previous
  observation implies that this element must be from $B \cap \Rstart$,
  from which it follows that the norm-filtering step runs for at most
  $|B \cap \Rstart| + 1$ iterations. This completes the proof of the first claim and hence proves the second
  claim as well.

  The final claim follows by the termination condition of the
  norm-filtering step.
\end{proof}

After the norm-filtering step,
$\alg$ computes $S_1$ and $S_2$ by calls to $\computeS$,
with both sets comprised of pairs of points with
large covariance-normalized inner product
(see Line~\ref{compS:loop-cond} of$\computeS$).
We require that $S_1$ and $S_2$ satisfy two key properties.
The first property is that $\scov{\cds}{S_\ell}$ upper bounds $\scov{\cds}{B \cap \Rnorm_\ell}$ in directions
where $\scov{\cds}{B \cap \Rnorm_\ell}$ is large.
We capture this property in Lemma~\ref{lem:bounded-bad-covariance-main} and
prove it in Appendix~\ref{sec:pf-bounded-bad-cov}.
The second property is that $\alg$ does not add too many uncontaminated points
to $S = S_1 \cup S_2$.
We capture this property in Lemma~\ref{lemma:few-good-points-in-S} and prove it
in Appendix~\ref{sec:pf-good-pts-in-S}.

\begin{lemma}
  \label{lem:bounded-bad-covariance-main}
  \ExecuteMetaData[\SectionsPath pf-sections/pf-bounded-bad-cov.tex]{bounded-bad-cov-lemma}
\end{lemma}

\begin{lemma}\label{lemma:few-good-points-in-S}
  \ExecuteMetaData[\SectionsPath pf-sections/pf-few-good-pts-in-S.tex]{few-good-pts-in-S-lemma}
\end{lemma}

We now briefly sketch the proof of Proposition~\ref{prop:single-iteration-result},
deferring the full proof to Appendix~\ref{sec:pf-prop-single-iteration}.
We use Lemma~\ref{lem:bounded-bad-covariance-main} to certify that $\norm{\Anorm \scov{\cds}{B \cap \Rnorm} \Anorm}$
is small if the condition in Line~\ref{alg:S-norm-check}
is satisfied, in which case $\alg$ terminates with output $\scov{\cds}{\Rnorm}$ satisfying
the first claim of the proposition.
If $\alg$ does not terminate, Lemmas~\ref{lem:bounded-bad-covariance-main} and \ref{lemma:few-good-points-in-S}
in combination with the operator norm resilience of $\uds$
ensure that the cumulative weight $\alg$ assigns to uncontaminated points in $S$
is smaller than the cumulative weight $\alg$ assigns to
the contaminated points in $S$ (Line~\ref{alg:w-def}).
This implies that Line~\ref{alg:random-filter} of $\alg$ will
filter more contaminated points than uncontaminated in expectation, thus proving the second claim of the proposition.

\subsection{Proof of Theorem~\ref{thm:alg-result}}\label{sec:thm-pf}
Define $G$ and $B$ as in the previous section.
Let $R_t$ denote the state of $R$ after the $t$-th iteration
of the outer while loop (and $R_0 = [n]$).
Let $M_t = |G \setminus R_t| + |R_t \setminus G|$ equal the number
removed uncontaminated points plus the number of remaining contaminated points
up to the $t$-th iteration.
Clearly $M_t \geq 0$ for all $t$ and $M_0 = |B| \leq \eps n$.
Let $T =
  \inf\{t \mid M_t > C\eps n \mbox{~or~}
  \alg~\mbox{terminates at iteration~} t\}$
and define
$N_t = M_{t \land T}$.

We first show $\{N_t\}$ is a supermartingale;
i.e., that $\E[N_t \mid N_0, \dots, N_{t-1}] \leq N_{t-1}$ for all $t \geq 0$.
Observe that the event $\{T \leq t-1\}$ and the random variables
$N_1, \dots, N_{t-1}$ are all $(R_1, \dots, R_{t-1})$-measurable.
We now condition on $R_1, \dots, R_{t-1}$.
If $T \leq t-1$, then clearly $N_t = N_{t-1}$, so there is nothing else to show.
Otherwise, we know that
$\alg$ has not terminated before the $t$-th iteration and also that
$M_{t-1} \leq C\eps n$, which implies
$|G \setminus R_{t-1}| \leq C\eps n$.
Observe that the change from $M_{t-1}$ to $M_t$ is equal to the
difference in how many
uncontaminated versus contaminated points were removed at the $t$-th iteration.
More precisely, because $B$ and $G$ partition $[n]$ and because
$R_t \subseteq R_{t-1}$, we have
\begin{align*}
  M_t-M_{t-1}
   & = |G \setminus R_t| + |R_t \setminus G| -
  (|G \setminus R_{t-1}| + |R_{t-1} \setminus G|) \\
   & = |G \cap (R_{t-1} \setminus R_t)|
  - |B \cap (R_{t-1} \setminus R_t)|.
\end{align*}
Because $|G \setminus R_{t-1}| \leq C\eps n$ and $M_t$ is conditionally independent
of $R_1,...,R_{t-2}$ given $R_{t-1}$,
the second claim of Proposition~\ref{prop:single-iteration-result} gives that
$\E[M_t-M_{t-1} \mid \xi_1, \dots, \xi_{t-1}] < 0$.
Then because $T > t-1$ in this case, we know that $N_{t-1} = M_{t-1}$ and
$N_t = M_t$.
Recalling that $N_1, \dots, N_{t-1}$ are measurable with respect to
$R_1, \dots, R_{t-1}$,
this proves that $\{N_t\}$ is a supermartingale.

Now because $\{N_t\}$ is a non-negative supermartingale with $N_0 \leq \eps n$
almost surely, we have
by Ville's inequality that with probability at least $1-C^{-1}$,
$\max_t N_t \leq C\eps n$.
This implies that with probability at least $1-C^{-1}$,
it holds that
$|G \setminus R_t| \leq C\eps n$ for all $t$ until $\alg$ terminates.
This means that the preconditions for
Proposition~\ref{prop:single-iteration-result}
hold until the $\alg$ terminates, at which point the first claim of
Proposition~\ref{prop:single-iteration-result} gives the desired claim that
the output $\SigmaAlg$ satisfies
\begin{align*}
  (1-\aop) \scov{\uds}{}
  \preceq \SigmaAlg
  \preceq \frac{1+\aop}{1-8\eps\tinner} \scov{\uds}{}.
\end{align*}

Finally, each iteration of $\alg$ removes at least one point from $R$.
Combined with the martingale analysis above, it follows for each iteration that $\alg$ removes at least one bad point
with probability at least $1/2$.
Thus, the expected number of iterations before $\alg$ terminates is at most $2 |B| \leq 2 \eps n$.
\section{Performance of \texorpdfstring{$\alg$}{ALG} for hypercontractive subgaussians}\label{sec:subg-perf}


Let $\cZ = \{Z_1, \dots, Z_n\} \simiid P$ where $P$ is $\sigma^2$-subgaussian (i.e.,
satisfies Definition~\ref{def:subg}), and let $\cX$ be an $\eps$-contamination of $\cZ$.
The following lemma gives the
parameters for which the dataset $\cZ$ is good.
We defer the proof
to Appendix~\ref{sec:lem-subg-goodness-pf}.
\begin{lemma}\label{lem:subg-goodness}
  Let $\delta > 0$,  $n \gtrsim \max\{\sigma^4 d + \log(1/\delta),
    \sigma^2(d+\log(n/\delta))\}$, and $\mop \leq n$.
  Then there exists $\bnorm, \binner, \aop$ satisfying
  \begin{align*}
    \bnorm^2 & = O(\sigma^2 \cdot (d + \log(n/\delta))) \\
    \binner  & = \sigma\sqrt{2\log(n^2/\delta)}\bnorm   \\
    \aop     &
    = 4\left(1 + 2\sigma^2\log\left(\frac{n}{2\mop}\right)\right) \cdot \frac{\mop}{n}
    + O\left(\sigma^2 \sqrt{\frac{d + \log(1/\delta)}{n}}\right)
  \end{align*}
  such that
  with probability at least $1-7\delta$,
  $\uds$ is $(\bnorm, 0, \binner, \mop, \aop)$-good.
\end{lemma}

Applying Theorem~\ref{thm:alg-result} with the parameters from Lemma~\ref{lem:subg-goodness} gives
the following performance guarantees of $\alg$ on hypercontractive
subgaussian data in Corollary~\ref{cor:subg-alg-perf}.

\begin{corollary}
  \label{cor:subg-alg-perf}
  Let $C > 0$ be a constant.
  Suppose $\log(n/\delta) \lesssim d$, $n \gtrsim \sigma^4 d$, and
  $\eps \lesssim \frac{\log n}{\sigma^2\sqrt{d}}$.
  Then there exists
  $\tnorm \asymp \sigma \sqrt{d}$ and
  $\tinner \asymp
    \sigma^2\log (n)\sqrt{d} \max\left\{1, \frac{1}{\eps \sqrt{n}}\right\}$
  such that with probability at least $1-7\delta-C$,
  $\SigmaAlg = \alg(\cds; \eps, \tnorm, \tinner)$
  satisfies
  \begin{align*}
    \err(\SigmaAlg, \scov{\uds}{})
    =
    O(\sigma^2\log n) \cdot (\eps\sqrt{d} + \sqrt{d/n}).
  \end{align*}
\end{corollary}
Corollary~\ref{cor:subg-alg-perf} bounds the error of $\alg$
with respect to the uncontaminated sample covariance $\scov{\uds}{}$.
One can combine this result with the concentration of the uncontaminated sample covariance
to the true population covariance (see Lemma~\ref{lem:subg-emp-cov-conc}) to obtain a formal
version of Theorem~\ref{thm:subg-result-informal}.
We defer the proof of Corollary~\ref{cor:subg-alg-perf} to Appendix~\ref{sec:subg-alg-perf-pf}.
\section{Performance of \texorpdfstring{$\alg$}{ALG} for moment bounded data}
\label{sec:heavy-perf}

\newcommand{\bclip}{b_{\mathsf{clip}}}


In the previous section, we were able to show show that the uncontaminated data $\cZ$ is good for parameters that imply
a near-linear sample complexity in $d$.
This is no longer the case when $\cZ$ is drawn i.i.d. from a moment-bounded distribution;
in particular, the heavy-tailed nature of these distributions means $\cZ$ will not be
$\bnorm$-norm bounded unless $\bnorm$ is polynomially larger than $d$.
In order then to achieve the near-linear sample complexity, we must consider
a clipped version of the distribution.

Let $\cZ = \{Z_1, \dots, Z_n\} \simiid P$ where $P$ has $k$-th moment bounded by $m_k$ (i.e.,
satisfies Definition~\ref{def:moment-bounded}), and let $\cX$ be an $\eps$-contamination of $\cZ$.
Let $\bclip > 0$ be a threshold and define
$\bar{z} = z\1{\norm{z}_\Sigma \leq \bclip}$. Let $\bar{Z} \sim \bar{P}$ for $Z \sim P$ with $\bar{\Sigma} = \E[\bar{Z}\bar{Z}^T]$,
and let $\bar{\uds} = \{\bar{Z}_1,...,\bar{Z}_n\}$.
Then by Lemma~\ref{lem:clipped-cov-err} we have
\begin{align}
  \dpsd(\bar{\Sigma},\Sigma)
  \leq m_k^k \left(\frac{\sqrt{d}}{\bclip}\right)^{k-2}, \label{eqn:heavy-tailed-display-1}
\end{align}
and by Lemma~\ref{lem:bdd-emp-cov-conc} we have
\begin{align}
  \dpsd(\scov{\bar{\uds}}{}, \bar{\Sigma}) = O\left(\sqrt{\frac{\bclip^2\log(d/\delta)}{n}}\right). \label{eqn:heavy-tailed-display-2}
\end{align}
Together, the above two displays imply that $\scov{\bar{\uds}}{}$ will be close to $\Sigma$
for $\bclip \gtrsim \sqrt{d}m_k^{\frac{k}{k-2}}$ and $n \gtrsim \bclip^2\log(d/\delta)$.
If $\cds$ was (hypothetically) an $\eps$-contamination of $\bar{\uds}$ instead of $\uds$,
we could then take the same approach as in the previous section by applying $\alg$ directly on $\cX$
to achieve the guarantee of Theorem~\ref{thm:heavy-tail-result-informal}.
Instead, we apply a separate preprocessing step (Algorithm~\ref{alg:preprocess}, or $\preprocess$) to $\cX$
before calling $\alg$.
Specifically,
we pick an initial norm threshold $\tclip \leq \frac{1}{2\sqrt{2}}\bclip$
and then apply $\alg$ to $\cY = \preprocess(\cX, \tclip)$.
\begin{algorithm}[h]
  \DontPrintSemicolon
  \caption{Preprocess step ($\preprocess$)}
  \label{alg:preprocess}
  \setcounter{AlgoLine}{0}
  \KwIn{Dataset $\cX = \{x_1, \dots, x_n\} \subset \R^d$,
    norm threshold $\tnorm$}
  \KwOut{Dataset $\{y_1,...,y_n\}$}
  \For{$i=1$ \KwTo $n$}{
    $y_i \leftarrow x_i \1{x_i^T \scovinv{\cds}{} x_i \leq \tnorm^2}$\;
  }
  \Return{$\{y_1,...,y_n\}$}
\end{algorithm}

The main technical challenge that arises with this approach is controlling the influence of the contaminated points on the preprocessing step.
If we set $\tnorm$ too large, then the contaminated points can conspire to prevent the preprocessing step from
removing many uncontaminated points whose norm with respect to $\Sigma$ is larger than $\bclip$, in which case
the remaining uncontaminated data may no longer be norm bounded (with suitable parameters);
but if we set $\tnorm$ too small then we risk removing too many uncontaminated points, in which case the remaining uncontaminated
data may no longer be inner product bounded.

We show in Lemma~\ref{lem:preproc-prop} that for $\tclip \asymp \bclip$ the preprocessing step
removes all but $\eps$-fraction of points whose norm with respect to $\Sigma$ is larger than $\bclip$.
We then show in Lemma~\ref{lem:heavy-goodness} that not only is $\bar{\cZ}$ itself good, but moreoever all sufficiently large
subsets of $\bar{\cZ}$ are also good.
We defer the proofs of these lemmas to Appendices~\ref{sec:preproc-prop} and~\ref{sec:pf-heavy-goodness}, respectively.

\begin{lemma}\label{lem:preproc-prop}
  Let
  $\bclip \gtrsim m_k^{\frac{k}{k-2}}\sqrt{d}$,
  $n \gtrsim \max\{\bclip^2\log(d/\delta), m_k^4(d+\log(1/\delta))\}$,
  and $\eps \lesssim m_k^{-\frac{2k}{k-2}}$.
  Let $\cY = \preprocess(\cX, \bclip/\sqrt{8})$.
  Then
  with probability at least
  \begin{align*}
    1-C_k m_k^k n^{1-k/2} - 2\delta - \exp\left(-\Omega\left( \left(\frac{8m_k\sqrt{d}}{\bclip}\right)^k n\right)\right),
  \end{align*}
  there exists $T \subseteq [n]$ satisfying
  $|T| \leq 2 \left(\frac{8 m_k \sqrt{d}}{\bclip}\right)^k n$
  and $\cY$ is a
  $\Big(\frac{9n\eps}{\bclip^2}\Big)$-contamination of
  $\bar{\uds}_{[n]\setminus T}$.
\end{lemma}

\begin{lemma}
  \label{lem:heavy-goodness}
  Let $\bclip \gtrsim m_k^{\frac{k}{k-2}}\sqrt{d}$ and
  \begin{align*}
    n \gtrsim \max\{\bclip^2\log(d/\delta), m_k^4(d+\log(1/\delta))\}.
  \end{align*}
  Let $\mop$ and $\mbig$ satisfy
  $\mop \lesssim m_k^{-\frac{2k}{k-2}}n$.
  Then there exists
  \begin{align*}
    \bnorm  & = O(\bclip)                                       \\ 
    \binner & = O(m_k n^{1/k} \bclip)                           \\ 
    \aop    & = O\left(\sqrt{\frac{\bclip^2 \log(d/\delta)}{n}}
    + m_k^2\sqrt{\frac{d + \log(1/\delta)}{n}}
    + m_k^2 \left(\frac{\mop}{n}\right)^{1-\frac{2}{k}}\right)
  \end{align*}
  such that with probability at least $1-O(\delta + n^{-\Omega(k\mbig)})$
  it holds for all $|\ipresset| \leq \min\{\mop,\mbig\}$ that
  $\bar{\uds}_{[n]\setminus \ipresset}$ is $(\bnorm, \mbig, \binner, \mop, \aop)$-good.
\end{lemma}

Together, Lemmas~\ref{lem:preproc-prop} and \ref{lem:heavy-goodness} imply there exists $\tclip$ such that the remaining uncontaminated data after the preprocessing step is good for parameters
that imply the result of Theorem~\ref{thm:heavy-tail-result-informal}. We formalize this claim in Corollary~\ref{cor:heavy-alg-perf}, which gives the main
performance guarantee of our estimator on moment-bounded data. We defer its proof to Appendix~\ref{sec:heavy-alg-perf-pf}.

\begin{corollary}\label{cor:heavy-alg-perf}
  Suppose $\bclip \gtrsim m_k^{\frac{k}{k-2}}\sqrt{d}$,
  $n \gtrsim \max\{\bclip^2\log(d/\delta), m_k^4(d+\log(1/\delta))\}$, and
  $\eps \lesssim m_k^{-\frac{2k}{k-2}}$.
  Let $\cY = \preprocess(\cX, \bclip/\sqrt{8})$.
  For any $\gamma > 0$ satisfying
  \begin{align*}
    \max\left\{\left(\frac{8 m_k \sqrt{d}}{\bclip}\right)^k, \frac{\eps n}{\bclip^2} \right\} \lesssim \gamma \lesssim m_k^{-\frac{2k}{k-2}},
  \end{align*}
  there exist parameters $\tnorm$ and $\tinner$ such that
  with probability at least
  \begin{align*}
    1 - C_k m_k^k n^{1-\frac{k}{2}} - \exp\left(-\Omega(\gamma n)\right)-O(\delta + n^{-\Omega(k\mbig)}),
  \end{align*}
  the output
  $\SigmaAlg = \alg(\cY, \frac{9n}{\bclip^2}\eps, \tnorm, \tinner)$ satisfies
  \begin{align*}
    \err(\SigmaAlg, \scov{\bar{\uds}}{})
     & \lesssim \log \bclip\sqrt{\frac{\bclip^2 \log(d/\delta)}{n}}
    + \frac{m_k \eps n^{1 + \frac{1}{k}}}{\bclip}
    + m_k^2 \gamma^{1-\frac{2}{k}}\log \bclip
    + \frac{\eps^2 n^2}{\bclip^2 \gamma}.
  \end{align*}
\end{corollary}

Corollary~\ref{cor:heavy-alg-perf} bounds the error of our estimator with respect to the sample covariance of the
clipped data.
Bounding the error with respect to $\Sigma$ then follows from combining the result of the corollary with the displays~\eqref{eqn:heavy-tailed-display-1}
and \eqref{eqn:heavy-tailed-display-2}.
For $m_k = O(1)$ and $n = \wt{\Omega}(d)$, taking
$\bclip = \wt{\Omega}(\sqrt{d})$
and $\gamma = \wt{O}(1)$
gives the result of Theorem~\ref{thm:heavy-tail-result-informal}.


\bibliography{bib}

\begin{thebibliography}{17}
\providecommand{\natexlab}[1]{#1}
\providecommand{\url}[1]{\texttt{#1}}
\expandafter\ifx\csname urlstyle\endcsname\relax
  \providecommand{\doi}[1]{doi: #1}\else
  \providecommand{\doi}{doi: \begingroup \urlstyle{rm}\Url}\fi

\bibitem[Cheng et~al.(2019)Cheng, Diakonikolas, Ge, and Woodruff]{ChengDiGeWo19}
Y.~Cheng, I.~Diakonikolas, R.~Ge, and D.~Woodruff.
\newblock Faster algorithms for high-dimensional robust covariance estimation.
\newblock \emph{arXiv:1906.04661 [cs.LG]}, 2019.

\bibitem[Cheng et~al.(2020)Cheng, Diakonikolas, Ge, and Soltanolkotabi]{ChengDiGeSo20}
Y.~Cheng, I.~Diakonikolas, R.~Ge, and M.~Soltanolkotabi.
\newblock High-dimensional robust mean estimation via gradient descent.
\newblock \emph{arXiv:2005.01378 [cs.LG]}, 2020.

\bibitem[Depersin and Lecu{\'{e}}(2019)]{DepersinLe19}
J.~Depersin and G.~Lecu{\'{e}}.
\newblock Robust subgaussian estimation of a mean vector in nearly linear time.
\newblock \emph{arXiv:1906.03058 [math.ST]}, 2019.

\bibitem[Diakonikolas and Kane(2022)]{DiakonikolasKa22}
I.~Diakonikolas and D.~M. Kane.
\newblock \emph{Algorithmic High-dimensional Robust Statistics}.
\newblock Cambridge University Press, 2022.

\bibitem[Diakonikolas et~al.(2016)Diakonikolas, Kamath, Kane, Li, Moitra, and Stewart]{DiakonikolasKaKaLiMoSt16}
I.~Diakonikolas, G.~Kamath, D.~M. Kane, J.~Li, A.~Moitra, and A.~Stewart.
\newblock Robust estimators in high dimensions without the computational intractability.
\newblock \emph{arXiv:1604.06443 [cs.DS]}, 2016.

\bibitem[Diakonikolas et~al.(2024)Diakonikolas, Hopkins, Pensia, and Tiegel]{DiakonikolasHoPeTi24b}
I.~Diakonikolas, S.~B. Hopkins, A.~Pensia, and S.~Tiegel.
\newblock {S}o{S} certificates for sparse singular values and their applications: Robust statistics, subspace distortion, and more.
\newblock \emph{arXiv:2412.21203 [cs.DS]}, 2024.

\bibitem[Dong et~al.(2019)Dong, Hopkins, and Li]{DongHoLi19}
Y.~Dong, S.~B. Hopkins, and J.~Li.
\newblock Quantum entropy scoring for fast robust mean estimation and improved outlier detection.
\newblock \emph{arXiv:1906.11366 [cs.DS]}, 2019.

\bibitem[Huber(1964)]{Huber64}
P.~J. Huber.
\newblock Robust estimation of a location parameter.
\newblock \emph{Annals of Mathematical Statistics}, 35\penalty0 (1):\penalty0 73--101, 1964.

\bibitem[Kothari et~al.(2018)Kothari, Steinhardt, and Steurer]{KothariStSt18}
P.~K. Kothari, J.~Steinhardt, and D.~Steurer.
\newblock Robust moment estimation and improved clustering via sum of squares.
\newblock In \emph{Proceedings of the Fiftieth Annual ACM Symposium on the Theory of Computing}, 2018.

\bibitem[Lai et~al.(2016)Lai, Rao, and Vempala]{LaiRaVe16}
K.~A. Lai, A.~B. Rao, and S.~Vempala.
\newblock Agnostic estimation of mean and covariance.
\newblock \emph{arXiv:1604.06968 [cs.DS]}, 2016.

\bibitem[Li and Ye(2020)]{LiYe20}
J.~Li and G.~Ye.
\newblock Robust gaussian covariance estimation in nearly-matrix multiplication time.
\newblock \emph{arXiv:2006.13312 [cs.DS]}, 2020.

\bibitem[Lugosi and Mendelson(2020)]{LugosiMe20}
G.~Lugosi and S.~Mendelson.
\newblock Multivariate mean estimation with direction-dependent accuracy.
\newblock \emph{arXiv:2010.11921 [math.ST]}, 2020.

\bibitem[Mendelson and Zhivotovskiy(2018)]{MendelsonZh18}
S.~Mendelson and N.~Zhivotovskiy.
\newblock Robust covariance estimation under ${L}_4-{L}_2$ norm equivalence.
\newblock \emph{arXiv:1809.10462 [math.ST]}, 2018.

\bibitem[Oliveira and Rico(2022)]{OliveiraRi22}
R.~I. Oliveira and Z.~F. Rico.
\newblock Improved covariance estimation: optimal robustness and sub-{G}aussian guarantees under heavy tails.
\newblock \emph{arXiv:2209.13485 [math.ST]}, 2022.

\bibitem[Tao(2012)]{Tao12}
T.~Tao.
\newblock \emph{Topics in Random Matrix Theory}, volume 132 of \emph{Graduate Studies in Mathematics}.
\newblock American Mathematical Society, 2012.

\bibitem[Tukey(1960)]{Tukey60}
J.~W. Tukey.
\newblock A survey of sampling from contaminated distributions.
\newblock In \emph{Contributions to Probability and Statistics: Essays in Honor of Harold Hotelling}, volume~2 of \emph{Stanford Studies in Mathematics and Statistics}. Stanford University Press, 1960.

\bibitem[Vershynin(2012)]{Vershynin12}
R.~Vershynin.
\newblock Introduction to the non-asymptotic analysis of random matrices.
\newblock In \emph{Compressed Sensing: Theory and Applications}, chapter~5, pages 210--268. Cambridge University Press, 2012.

\end{thebibliography}
\bibliographystyle{abbrvnat}

\newpage
\appendix

\section{Helper Lemmas}\label{sec:helper-lemmas}
\subsection{Basic Lemmas}
\begin{lemma}[Properties of $\dpsd$]\label{lem:dpsd-prop}
  Let $A, B$ be p.s.d. matrices such that $\dpsd(A, B) \leq c < 1$.
  Then,
  \begin{align*}
    \norm{B^{1/2}A^{\dag/2}} \leq \sqrt{1+c}
    ~~~\mbox{and}~~~
    \dpsd(B, A) \leq \frac{c}{1-c}
  \end{align*}
\end{lemma}
\begin{proof}
  First, suppose for contradiction that $\colspace(A) \neq \colspace(B)$.
  Then because $\colspace(B) \subseteq \colspace(A)$,
  we can pick a normal vector $v \in \colspace(A)$ that is orthogonal to
  $\colspace(B)$ and moreover, there exists a unique vector $u \in \colspace(A)$
  such that $A^{\dag/2}u = v$.
  Then we have that
  \begin{align*}
    u^TA^{\dag/2}(B - A)A^{\dag/2}u
    = v^TBv - u^TA^{\dag/2}AA^{\dag/2}u
    = u^Tu,
  \end{align*}
  which implies that $\norm{A^{\dag/2}(B - A)A^{\dag/2}} \geq 1$, contradicting
  our assumption $\dpsd(A, B) < 1$.
  Thus, $\colspace(A) = \colspace(B)$ and so by restricting to vectors
  in $\colspace(A)$, we can assume without loss of generality that $A, B$ are
  both invertible.
  Then, because $\norm{A^{-1/2}BA^{-1/2} - I} = \dpsd(A, B) \leq c$,
  we know that the spectrum $A^{-1/2}BA^{-1/2}$ is contained in $[1-c, 1+c]$.
  Because for any matrix $M$, the singular values of $MM^T$ are the square of
  the singular values of $M$, we have that the spectrum of
  $B^{1/2}A^{-1/2}$ is contained in $[\sqrt{1-c}, \sqrt{1+c}]$,
  proving the first claim.
  We then have that inverse
  $A^{1/2}B^{-1/2}$ has spectrum contained in $[1/\sqrt{1+c}, 1/\sqrt{1-c}]$,
  which implies the spectrum of $B^{-1/2}(A-B)B^{-1/2}$
  is contained in $[\frac{1}{1+c} - 1, \frac{1}{1-c}-1]$, proving the second claim.
\end{proof}
\subsection{Subgaussian concentration lemmas}
Throughout this section,
we let $Z_1, \dots, Z_n \simiid P$ for a $\sigma^2$-hypercontractive subgaussian distribution $P$ over $\R^d$ with mean zero
and covariance $\Sigma$ (Definition~\ref{def:hyper-subg}).

\begin{definition}[Subgaussian random variable]\label{def:subg}
  A random variable $Y$ in $\R$ is $\sigma^2$-subgaussian if for all
  $\lambda \in \R$, $\E[e^{\lambda(Y-\E Y)}] \leq e^{\frac{\sigma^2 \lambda^2}{2}}$.
\end{definition}

Note for $Z \sim P$ and any $v \in \S^{d-1}$ that
$v^T\Sigma^{\dag/2} Z$ is $\sigma^2$-subgaussian.

\begin{definition}[Sub-exponential random variable]\label{def:subexp}
  A random variable $Y$ in $\R$ is $\nu$-sub-exponential if for all
  $\lambda \in [-1/\nu, 1/\nu]$, $\E[e^{\lambda(Y-\E Y)}] \leq e^{\nu^2 \lambda^2}$.
\end{definition}

\begin{lemma}[Tail bounds of subgaussians {\citep[Lemma 5.5]{Vershynin12}}]
  \label{lem:subg-tail-bd}
  Let $c > 0$ be a sufficiently small constant.
  Let $Y$ be a mean zero $\sigma^2$-subgaussian random variable in $\R$.
  Then, for all $t \geq 0$, $\P(|Y| > t) \leq \exp(1 - c\frac{t^2}{\sigma^2})$.
\end{lemma}

\begin{lemma}[Concentration of empirical covariance
      {\citep[Theorem 5.39]{Vershynin12}}]\label{lem:subg-emp-cov-conc}
  Let $\delta > 0$ and suppose $n \gtrsim \sigma^4 d + \log(1/\delta)$.
  Then with probability at least $1-\delta$,
  \begin{align*}
    \dpsd\left(\frac{1}{n}\sum_{i=1}^n Z_i Z_i^T,\Sigma\right)
    = O\left(\sigma^2 \sqrt{\frac{d}{n}} + \sqrt{\frac{\log(1/\delta)}{n}}\right).
  \end{align*}
\end{lemma}
\begin{proof}
  The result follows almost immediately from applying \citet[Theorem 5.39]{Vershynin12} to the random variables $\Sigma^{\dag/2} Z_1,...,\Sigma^{\dag/2}Z_n$.
\end{proof}

\begin{lemma}[Bernstein-type inequality {\citep[Proposition 5.16]{Vershynin12}}]
  \label{lem:bernstein}
  Let $c > 0$ be a sufficiently small universal constant.
  Let $Y_1, \dots, Y_n$ be independent $\nu$-sub-exponential random variables.
  Then for any vector $(a_1, \dots, a_n) \in \R^n$ and every $t \geq 0$,
  \begin{align*}
    \P\left(
    \left| \sum_{i=1}^n a_i Y_i \right| \geq t
    \right)
    \leq 2\exp\left[-c \min\left\{\frac{t^2}{\nu^2 \norm{a}_2^2},
      \frac{t}{\nu\norm{a}_\infty}\right\}\right].
  \end{align*}
\end{lemma}

\begin{lemma}[Square of subgaussian is sub-exponential {\citep[Lemma 5.14]{Vershynin12}}]
  \label{lem:sq-subg-is-sube}
  If $Y$ is $\sigma^2$-subgaussian, then $Y^2$ is $2\sigma^2$-sub-exponential.
\end{lemma}

\begin{lemma}[Subgaussian $\ell_2$-norm boundedness]\label{lem:subg-norm-boundedness}
  Let $\delta > 0$.
  With probability at least $1-\delta$,
  \begin{align*}
    \sup_{i \in [n]} \norm{Z_i}_\Sigma^2
    \leq \sigma^2 \cdot O(d + \log(n/\delta)).
  \end{align*}
\end{lemma}
\begin{proof}
  %
  We first observe that
  $\norm{Z_i}_\Sigma = \norm{\Sigma^{\dag/2}Z_i}$.
  Recalling that $v^T\Sigma^{\dag/2} Z_i$ is $\sigma^2$-subgaussian for any
  $v \in \S^{d-1}$, we have by a standard packing argument
  (see, e.g., \citet[Chapter 5]{Vershynin12}) and Lemma~\ref{lem:subg-tail-bd} that
  $
    \P(\norm{\Sigma^{\dag/2}Z_i} > t)
    \leq 4^d \exp(1-ct^2/\sigma^2)
  $.
  Picking $t^2 \gtrsim \sigma(d + \log(n/\delta))$, we have by a union bound
  over $i \in [n]$
  that with probability at least $1-\delta$,
  $\sup_{i \in [n]} \norm{Z_i}_\Sigma^2 \leq t^2$, proving the claim.
\end{proof}
\subsection{Moment bounded concentration lemmas}
\label{sec:heavy-helpers}

Throughout this section,
we let $Z_1, \dots, Z_n \simiid P$ for a distribution $P$ over $\R^d$ with mean zero,
covariance $\Sigma$, and $k$-th moments bounded by $m_k$ (Definition~\ref{def:moment-bounded}).

\begin{definition}[Moment bounded random variable]
  A random variable $Y \in \R$ has $k$-th moments bounded by $m_k$ if
  $\E [|Y|^k]^{1/k} \leq m_k$.
\end{definition}

Note for $Z \sim P$ and any $v \in \S^{d-1}$ that $v^T\Sigma^{\dag/2} Z$ has $k$-th moments bounded by $m_k$.

\begin{lemma}\label{lem:kmom-avg-tail-bd}
  Let $Y_1, \dots Y_n \in \R$ be independent random variables with mean zero
  and $k$-th moment bounded by $m_k$. Then
  \begin{align*}
    \P\left(\frac{1}{n}\sum_{i=1}^n Y_i \geq \alpha \right)
     & \leq 2\left(\sqrt{\frac{ek}{2}}\cdot \frac{m_k}{\alpha\sqrt{n}}\right)^{k}.
  \end{align*}
\end{lemma}
\begin{proof}
  We adapt an argument from \citet[Section 2.1]{Tao12} for bounding the deviation
  of the mean of $k$-moment bounded random variables.
  \begin{align*}
    \P\left(\frac{1}{n}\sum_{i=1}^n Y_i \geq \alpha \right)
     & \leq \P\left(\bigg|\sum_{i=1}^n Y_i\bigg|^k \geq (n\alpha)^k \right)     \\
     & \leq \frac{1}{(n\alpha)^k}\E\left[\bigg|\sum_{i=1}^n Y_i\bigg|^k\right].
  \end{align*}
  Expanding the term in the expectation, we have
  \begin{align*}
    \E\left[\bigg|\sum_{i=1}^n Y_i\bigg|^k\right]
     & = \sum_{1 \leq i_1, \dots, i_k \leq n} \E[Y_{i_1} \dots Y_{i_k}].
  \end{align*}
  Notice that for any choice of $i_1, \dots, i_k$ such that any specific value
  occurs exactly once, the expectation is $0$ by independence and the mean zero
  assumption.
  For other choices of $i_1, \dots, i_k$, the expectation is bounded by
  $m_k^k$ by independence and Jensen's inequality.
  Thus, we count the number of ways to choose $i_1, \dots, i_k$ such that
  no value appears exactly once.
  For such $i_1, \dots, i_k$, note that only at most $k/2$ distinct values
  can occur.
  Let $N_r$ be the number of ways to assign $i_1, \dots, i_k$ such that
  exactly $k/2-r$ values appear.
  Then, following \citet[Section 2.1]{Tao12},
  $N_r \leq \choose{n}{\frac{k}{2}-r}\left(\frac{k}{2}-r\right)^k
    \leq (en)^{\frac{k}{2} - r}\left(\frac{k}{2}\right)^{k/2+r}$,
  by Stirling's formula..
  Thus the total number of assignments is bounded by
  \begin{align*}
    \sum_{r=0}^{k/2} N_r
    \leq \left(\frac{ekn}{2}\right)^{k/2}
    \sum_{r=0}^{k/2} \left(\frac{k}{2en}\right)^r
    \leq \left(\frac{ekn}{2}\right)^{k/2} \left(\frac{1}{1-\frac{k}{2en}}\right)
    \leq 2 \left(\frac{ekn}{2}\right)^{k/2}.
  \end{align*}
  Thus,
  \begin{align*}
    \P\left(\frac{1}{n}\sum_{i=1}^n Y_i \geq \alpha \right)
     & \leq 2\left(\sqrt{\frac{ek}{2}}\cdot \frac{m_k}{\alpha\sqrt{n}}\right)^{k}.
  \end{align*}
\end{proof}

\begin{lemma}
  \label{lem:heavy-norm-tail-bd}
  Suppose $k \geq 2$. Then for $Z \sim P$ we have
  \begin{align*}
    \E[\norm{Z}_\Sigma^k] \leq m_k^k d^{k/2}
    ~~~\textrm{and}~~~
    \P(\norm{Z}_\Sigma > M)
    \leq \left(\frac{m_k \sqrt{d}}{M}\right)^k.
  \end{align*}
\end{lemma}
\begin{proof}
  The last claim follows from the first and Markov's inequality, and so we
  turn to proving the first claim.
  We have,
  \begin{align*}
    \E[\norm{Z}_\Sigma^k]
     & = \E\left[\left(\sum_{j=1}^d (\Sigma^{\dag/2}Z \cdot e_j)^2\right)^{k/2}\right]                      \\
     & = d^{k/2} \E\left[\left( \frac{1}{d} \sum_{j=1}^d (\Sigma^{\dag/2}Z \cdot e_j)^2\right)^{k/2}\right] \\
     & \leq d^{k/2} \E\left[\frac{1}{d} \sum_{j=1}^d |\Sigma^{\dag/2} Z \cdot e_j|^k\right]                 \\
     & \leq m_k^k d^{k/2},
  \end{align*}
  where the first inequality is from Jensen's inequality and the assumption that
  $k \geq 2$ and the second inequality follows from the moment boundedness
  assumption.
\end{proof}

\begin{lemma}\label{lem:clipped-cov-err}
  Let $Z \sim P$ and $t > 0$. Let
  $\bar{Z} = Z\1{\norm{Z}_{\Sigma} \leq t}$ and let
  $\bar{\mu} = \E[\bar{Z}]$ and
  $\bar{\Sigma} = \E[\bar{Z}\bar{Z}^T]$. Then,
  \begin{align*}
    \norm{\bar{\mu}}_\Sigma \leq m_k^k \left(\frac{\sqrt{d}}{t}\right)^{k-1}
    ~~~\mbox{and}~~~
    \dpsd(\bar{\Sigma}, \Sigma)
    \leq m_k^k \left(\frac{\sqrt{d}}{t}\right)^{k-2}.
  \end{align*}
\end{lemma}
\begin{proof}
  To bound $\norm{\bar{\mu}}_\Sigma$, recall that $P$ has mean zero and so
  \begin{align*}
    \norm{\bar{\mu}}_\Sigma
     & = \norm{\E[Z - Z\1{\norm{Z}_\Sigma \leq t}]}_\Sigma                            \\
     & = \norm{\E[Z\1{\norm{Z}_\Sigma > t}]}_\Sigma                                   \\
     & = \sup_{v \in \S^{d-1}}(\Sigma^{\dag/2} \E[Z\1{\norm{Z}_\Sigma > t}] \cdot v).
  \end{align*}
  For every $v \in \S^{d-1}$, we have by H\"{o}lder's inequality
  and Lemma~\ref{lem:heavy-norm-tail-bd}
  that
  \begin{align*}
    \E[(\Sigma^{\dag/2} Z\cdot v)\1{\norm{Z}_\Sigma > t}]
     & \leq \E[|\Sigma^{\dag/2}Z \cdot v|^k]^{\frac{1}{k}}
    \P(\norm{Z}_\Sigma > t)^{1-\frac{1}{k}}
    \leq m_k^k \left(\frac{\sqrt{d}}{t}\right)^{k-1}.
  \end{align*}
  We now turn to the covariance.
  Trivially, we have $\colspace(\bar{\Sigma}) \subseteq \colspace(\Sigma)$ and so
  $\dpsd(\bar{\Sigma}, \Sigma) < \infty$.
  Thus, it suffices to bound $v^T \Sigma^{\dag/2}(\Sigma - \bar{\Sigma}) \Sigma^{\dag/2}v$ for each $v \in \S^{d-1} \cap \colspace(\Sigma)$.
  To this end, we have by H\"{o}lder's inequality that
  \begin{align*}
    v^T\Sigma^{\dag/2}(\Sigma-\bar{\Sigma})\Sigma^{\dag/2}v
     & = \E[(\Sigma^{\dag/2}Z \cdot v)^2 - (\Sigma^{\dag/2}\bar{Z} \cdot v)^2] \\
     & = \E[(\Sigma^{\dag/2}Z \cdot v)^2 \1{\norm{Z}_\Sigma > t}]              \\
     & \leq \E[|v^T\Sigma^{\dag/2}Z|^k]^{\frac{2}{k}}
    \P(\norm{Z}_\Sigma > t)^{1 - \frac{2}{k}}.
  \end{align*}
  The moment boundedness assumption gives
  $\E[|v^T\Sigma^{\dag/2}Z|^k] \leq m_k^k$ and
  Lemma~\ref{lem:heavy-norm-tail-bd} gives
  $\P(\norm{Z}_\Sigma > t) \leq \left(\frac{m_k \sqrt{d}}{t}\right)^k$,
  proving the claim.
\end{proof}

\begin{lemma}[Large points have large sample covariance norms]
  \label{lem:heavy-large-points}
  Let $C_k$ be a sufficiently large constant that depends only on $k$.
  Let $t > 0$ and $n \geq 2t^2$.
  With probability at least $1-C_km_k^k n^{1-k/2}$,
  we have for all $i \in [n]$ that if $\norm{Z_i}_\Sigma > 2t$, then
  $Z_i^T \scovinv{\uds}{} Z_i > t^2$.
\end{lemma}
\begin{proof}
  Almost surely $\spn \cZ \subseteq \colspace(\Sigma)$,
  so we can assume
  both $Z_i^T\Sigma^{\dag}Z_i$ and
  $Z_i^T \scovinv{\uds}{} Z_i$ are finite.
  We show for each $i \in [n]$ that
  $\norm{Z_i}_\Sigma > 2t$ implies $Z_i^T \scovinv{\uds}{} Z_i > t^2$
  with probability at least $1-C_k m_k n^{-k/2}$.
  The desired claim then follows via a union bound over $i \in [n]$.

  Applying the Sherman-Morrison formula gives
  \begin{align*}
    Z_i^T \scovinv{\uds}{} Z_i
     & = Z_i^T \left(\scov{\uds}{-i} + \frac{1}{n} Z_i Z_i^T\right)^{\dag} Z_i \\
     & = Z_i^T\scovinv{\uds}{-i}Z_i
    - \frac{(Z_i^T\scovinv{\uds}{-i}Z_i)^2}{n + Z_i^T\scovinv{\uds}{-i}Z_i}    \\
     & = \frac{n}{n + Z_i^T\scovinv{\uds}{-i}Z_i}
    \cdot Z_i^T\scovinv{\uds}{-i}Z_i.
  \end{align*}
  The final quantity is increasing in $Z_i^T\scovinv{\uds}{-i}Z_i$, so
  if $Z_i^T\scovinv{\uds}{-i}Z_i > 2t^2$ then
  \begin{align*}
    Z_i^T \scovinv{\uds}{} Z_i
    > \frac{2t^2 n}{n + 2t^2}
    \geq t^2.
  \end{align*}
  Thus, it suffices to show that $\norm{Z_i}_\Sigma > 2t$ implies
  $Z_i^T\scovinv{\uds}{-i}Z_i > 2t^2$ with probability at least
  $1-C_k m_k^k n^{-k/2}$. To this end,
  we have by Cauchy-Schwarz that
  \begin{align*}
    \norm{Z_i}_\Sigma^4
    = (Z_i^T \Sigma^\dag Z_i)^2
    = (Z_i^T \scovisqrt{\uds}{-i} \scovsqrt{\uds}{-i}\Sigma^\dag Z_i)^2
    \leq (Z_i^T \scovinv{\uds}{-i} Z_i) \cdot (Z_i^T \Sigma^\dag \scov{\uds}{-i} \Sigma^\dag Z_i),
  \end{align*}
  which implies for $V_i = \frac{\Sigma^{\dag/2}Z_i}{\norm{Z_i}_\Sigma}$ that
  \begin{align*}
    Z_i^T \scovinv{\uds}{-i}Z_i
    \geq \norm{Z_i}_\Sigma^2 (V_i^T \Sigma^{\dag/2} \scov{\uds}{-i} \Sigma^{\dag/2} V_i)^{-1}.
  \end{align*}
  Because $V_i$ and $\cZ_{-i}$ are independent, we can fix $V_i = v$ and
  show a high probability upper bound on $v^T \Sigma^{\dag/2} \scov{\uds}{-i} \Sigma^{\dag/2} v$ over the
  distribution of $\cZ_{-i}$.
  Expanding gives
  \begin{align*}
    v^T \Sigma^{\dag/2} \scov{\uds}{-i} \Sigma^{\dag/2} v
     & = \frac{1}{n}\sum_{j \neq i} (\Sigma^{\dag/2}Z_j \cdot v)^2,
  \end{align*}
  where by Lemma~\ref{lem:kmom-avg-tail-bd} we have
  \begin{align*}
    \P\left(\frac{1}{n-1}\sum_{j \neq i} (\Sigma^{\dag/2}Z_j \cdot v)^2 - 1 \geq \alpha \right)
     & \leq 2\left(\sqrt{\frac{ek}{2}}\cdot \frac{m_k}{\alpha\sqrt{n-1}}\right)^{k}.
  \end{align*}

  Thus, with probability at least $1-C_k m_k n^{-k/2}$ over the randomness
  of $\cZ_{-i}$, we have that $v^T \Sigma^{\dag/2} \scov{\uds}{-i} \Sigma^{\dag/2} v \leq 2$.
  Therefore, with at least that same probability we have
  \begin{align*}
    Z_i^T \scovinv{\uds}{-i}Z_i \geq \norm{Z_i}_\Sigma^2/2,
  \end{align*}
  from which it follows that $\norm{Z_i}_\Sigma > 2t$ implies
  $Z_i^T \scovinv{\uds}{-i}Z_i > (2t)^2/2 \geq 2t^2$.
\end{proof}

\begin{lemma}[Sample covariance of bounded random variables]
  \label{lem:bdd-emp-cov-conc}
  Let $Y_1, \dots, Y_n \iid P$ such that
  $\E[Y_1Y_1^T] = \Sigma$ and $\norm{Y_1}_\Sigma \leq M$ almost surely.
  Then with probability at least $1-\delta$ we have
  \begin{align*}
    \dpsd\left(\frac{1}{n}\sum_{i=1}^n Y_i Y_i^T,\Sigma\right)
    = O\left(\sqrt{\frac{M^2 \log(d/\delta)}{n}}\right).
  \end{align*}
\end{lemma}
\begin{proof}
  This follows immediately from applying \citet[Theorem 5.41]{Vershynin12}
  on the whitened data $\Sigma^{\dag/2}Y_i$.
  Note that in our setting $\E[\Sigma^{\dag/2}Y_1]$ need not be $0$,
  but \citet[Theorem 5.41]{Vershynin12} only states their
  concentration result for isotropic random variables.
  This is not a problem because the proof of \citet[Theorem 5.41]{Vershynin12}
  does not rely on the mean being zero, it only requires that the second moment
  be the identity.
\end{proof}

\begin{lemma}[{\citet[Proposition 3.4]{OliveiraRi22}}]\label{lem:oliveira-prop}
  Suppose $Y_1, \dots, Y_n$ are i.i.d. nonnegative real random variables with
  $\E[Y_1^q] < \infty$ for $q \geq 2$ and let $W_1, \dots, W_n$ be an
  $\eta$-contamination of $Y_1, \dots, Y_n$.
  Let $c > 0$ be a universal constant, let $\alpha \in (0, 1)$.
  For $k = \lfloor \eta n\rfloor + \lceil c\eta n + \log(2/\alpha)\rceil$,
  let $T_k = \inf_{S \subseteq [n], |S|=k} \sum_{i \in S} \frac{W_i}{n-k}$.
  If $k < n$, then
  \begin{align*}
    \P\left(|T_k - \E[Y_1]|
    \leq C\sqrt{\frac{\E[Y_1^2]\log(2/\alpha)}{n}}
    + C_c\E[Y_1^q]^{\frac{1}{q}}\eta^{1-\frac{1}{q}}\right)
    \geq 1-\alpha,
  \end{align*}
  for universal constants $C, C_c > 0$ where $C_c$ only depends on $c$.
\end{lemma}
\section{Proofs for Section~\ref{sec:single-iteration-analysis}}
  \label{sec:pf-sec-single-iteration}
\subsection{Proof of Proposition~\ref{prop:single-iteration-result}}
  \label{sec:pf-prop-single-iteration}
Rather than using Lemma~\ref{lem:bounded-bad-covariance-main} directly, we
will instead use the following corollary, which converts the bound in
Lemma~\ref{lem:bounded-bad-covariance-main} from being in terms of
$\Rnorm_\ell$ and $S_\ell$ to being in terms of $\Rnorm$ and $S$.

\begin{corollary}[Corollary of Lemma~\ref{lem:bounded-bad-covariance-main}] \label{lem:bounded-bad-covariance}
  Suppose $m \geq \frac{12e \tnorm^2\log(\tnorm)}{\tinner}$.
  For all $T \subseteq \Rnorm$ such that $|T| \leq m$ and all $v \in \S^{d-1}$, we have
  \begin{align*}
    & v^T\Anorm\scov{\cds}{T}\Anorm v \\
    &\qquad \leq
    \max\left\{ 
    \frac{8m}{n}\tinner,
    32e \log(\tnorm)
    \cdot v^T \Anorm\scov{\cds}{S}\Anorm v
    \right\}.
  \end{align*}
\end{corollary}
\begin{proof-of-corollary}[\ref{lem:bounded-bad-covariance}]
  For $\ell \in \{1,2\}$, applying Lemma~\ref{lem:bounded-bad-covariance-main} to
  the subset $T \cap \Rnorm_\ell$ gives for all $u \in \R^d$
  \begin{align*}
    u^T\scov{\cds}{T \cap \Rnorm_\ell} u
    &\leq
    \max\left\{
    \frac{4m}{n}\tinner \cdot u^T \scov{\cds}{\Rnorm_{\ell'}} u,
    16e \log(\tnorm)
    \cdot u^T \scov{\cds}{S_\ell} u
    \right\} \\
    &\leq
    \max\left\{
    \frac{4m}{n}\tinner \cdot u^T \scov{\cds}{\Rnorm} u,
    16e \log(\tnorm)
    \cdot u^T \scov{\cds}{S} u
    \right\},
  \end{align*}
  where the second inequality is because $\scov{\cds}{\Rnorm_{\ell'}} \preceq \scov{\cds}{\Rnorm}$
  and $\scov{\cds}{S_\ell} \preceq \scov{\cds}{S}$.

  Then we have that
  \begin{align*}
    u^T\scov{\cds}{T} u
    &= u^T\scov{\cds}{T \cap \Rnorm_1} u + u^T\scov{\cds}{T \cap \Rnorm_2} u \\
    &\leq
    2\max\left\{
      \frac{4m}{n}\tinner \cdot u^T \scov{\cds}{\Rnorm} u,
      16e \log(\tnorm)
      \cdot u^T \scov{\cds}{S} u
      \right\},
  \end{align*}
  from which the desired claim follows by letting $u = \Anorm v$.
\end{proof-of-corollary}

We now prove Proposition~\ref{prop:single-iteration-result}.

First consider the case that
$\norm{\Anorm \scov{\cds}{S} \Anorm}
  \leq \frac{\eps \tinner}{4e \log(\tnorm)}$.
By Line~\ref{alg:S-norm-check}, $\alg$ immediately terminates
and outputs $\SigmaAlg = \scov{\cds}{\Rnorm}$.
Thus, we need to show that in this case,
$(1-\aop) \cdot \scov{\uds}{}
  \preceq \scov{\cds}{\Rnorm}
  \preceq \frac{1+\aop}{1-8\eps\tinner} \cdot \scov{\uds}{}$.
Because we assumed $n \geq \frac{12e\tnorm^2\log(\tnorm)}{\eps\tinner}$,
we can apply Corollary~\ref{lem:bounded-bad-covariance} with $m = \eps n$
and $T = B \cap \Rnorm$ to obtain
\begin{align*}
  &\norm{\Anorm\scov{\cds}{B \cap \Rnorm}\Anorm} \\
  &\qquad\leq
  \max\left\{
  8\eps\tinner,
  32 e \log(\tnorm)
  \norm{\Anorm\scov{\cds}{S}\Anorm}
  \right\} \\
  &\qquad\leq 8\eps\tinner.
\end{align*}
Then, because
$\scov{\cds}{B \cap \Rnorm} = \scov{\cds}{\Rnorm} - \scov{\cds}{G \cap \Rnorm}$,
we have $\dpsd(\scov{\cds}{G \cap \Rnorm}, \scov{\cds}{\Rnorm}) \leq 8\eps\tinner$,
which implies
$\scov{\cds}{\Rnorm} \preceq (1-8\eps\tinner)^{-1}\scov{\cds}{G \cap \Rnorm}$.
We also have $\scov{\cds}{G \cap \Rnorm} \preceq \scov{\cds}{\Rnorm}$ because
$G \cap \Rnorm \subseteq \Rnorm$.
Furthermore, Lemma~\ref{lem:norm-filter-props}
gives that $|G \cap \Rnorm| = |G \cap \Rstart| \geq (1-(1 + C)\eps) n \geq n - \mop$,
and so operator-norm resilience gives by Lemma~\ref{lem:sub-contam-bounded-cov}
that
$(1-\aop)\scov{\uds}{}
\preceq \scov{\cds}{G \cap \Rnorm}
\preceq (1+\aop)\scov{\uds}{}$.
Combining all these bounds yields,
\begin{align*}
  (1-\aop)\scov{\uds}{}
  \preceq \scov{\cds}{\Rnorm}
  \preceq \frac{1+\aop}{1-8\eps\tinner}\scov{\uds}{},
\end{align*}
which proves the desired claim for this case.

Now consider the case that
$\norm{\Anorm \scov{\cds}{S} \Anorm}
  > \frac{\eps \tinner}{4e \log(\tnorm)}$.
In this case, $\alg$ does not terminate at the current iteration,
and so we need to show that $\alg$ will remove more points from $B$ than from
$G$ on average.
Observe that $\alg$ only removes points in the norm filter
of Line~\ref{alg:norm-filter-pt-removal}
and the randomized filter of Line~\ref{alg:random-filter}.
Moreover, Lemma~\ref{lem:norm-filter-props} gives that the norm filter only
removes points in $B$.
Thus, it suffices to show that the randomized filter will on average remove more 
points from $B$ than from $G$; i.e., that
$\E[|B \cap (\Rnorm \setminus \Rend)|] > \E[|G \cap (\Rnorm \setminus \Rend)|]$.
Because $B$ and $G$ partition $[n]$, this claim is equivalent to
\begin{align*}
  \E[|G \cap (\Rnorm \setminus \Rend)|]
  < \frac{1}{2}\E[|\Rnorm \setminus \Rend|].
\end{align*}
To prove this, we first observe that after
the if-statement check in Line~\ref{alg:S-norm-check} fails,
$\alg$ computes $v_S \in \S^{d-1}$ such that
\begin{align*}
  v_S^T\Anorm \scov{\cds}{S} \Anorm v_S
  = \norm{\Anorm \scov{\cds}{S} \Anorm}
  > \frac{\eps \tinner}{4e \log(\tnorm)}.
\end{align*}
Then, $\alg$ filters each point $i \in \Rnorm$ with probabilty
$\frac{1}{\wmax} (x_i \cdot \Anorm v_S)^2\1{i \in S}$.
By linearity of expectation, we thus have
\begin{align}
  \E[|\Rnorm \setminus \Rend|]
   & = \frac{1}{\wmax} \sum_{i \in \Rnorm} (x_i \cdot \Anorm v_S)^2\1{i \in S}
    \nonumber \\
   & = \frac{1}{\wmax}\sum_{i \in S} v_S^T \Anorm x_ix_i^T \Anorm v_S
     \nonumber \\
   & = \frac{n}{\wmax} \cdot v_S^T \Anorm \scov{\cds}{S} \Anorm v_S
     \nonumber \\
   & > \frac{\eps \tinner}{4e \log(\tnorm)} \cdot \frac{n}{\wmax}
     \label{eq:total-removed-lb}.
\end{align}

On the other hand, we can bound the expected number of good points removed by
\begin{align}
  \E[|G \cap (\Rnorm \setminus \Rend)|]
   & = \frac{1}{\wmax} \sum_{i \in G \cap \Rnorm} (x_i \cdot \Anorm v_S)^2\1{i \in S} 
    \nonumber \\
   & = \frac{n}{\wmax} \cdot v_S^T \Anorm \scov{\cds}{G \cap S} \Anorm v_S 
    \nonumber \\
   & \leq \frac{n}{\wmax} \norm{\Anorm \scov{\cds}{G \cap S}\Anorm}
    \label{eq:removed-good-ub}.
\end{align}
To bound $\norm{\Anorm \scov{\cds}{G \cap S}\Anorm}$, we begin by bounding the
number of good points added to $S$.
First note that
$|(G \cap \Rnorm)| = n - |B| - |G \setminus \Rstart|
\geq n - (1+C)\eps n$.
Then because $\eps \leq \frac{1}{4(1+C)}$ and $\mop \geq (1+C)\eps n$,
we may apply Lemma~\ref{lemma:few-good-points-in-S} to obtain
\begin{align*}
  |G \cap S|
  &\leq 4\mbig + |B \cap \Rnorm|
    + \frac{72\tnorm^2}{\tinner}(n - |G \cap \Rnorm|).
\end{align*}
Because we also assume that $\mop \geq 4\mbig + \left(1 + \frac{72(1+C)\tnorm^2}{\tinner}\right)\eps n$,
the above display implies $|G \cap S| \leq \mop$.
Thus, $(\mop, \aop)$-operator norm resilience implies
$\scov{\cds}{G \cap S} \preceq \aop \scov{\uds}{}$ and so
\begin{align*}
  \norm{\Anorm\scov{\cds}{G \cap S}\Anorm}
   & \leq \aop \norm{\Anorm\scov{\uds}{}\Anorm}                                          \\
   & = \aop \norm{\scovsqrt{\uds}{}\scovinv{\cds}{\Rnorm}\scovsqrt{\uds}{}}.
\end{align*}
Then observe that
$\scovinv{\cds}{\Rnorm} \preceq \scovinv{\cds}{G \cap \Rnorm}
= \scovinv{\uds}{G \cap \Rnorm} \preceq (1-\aop)^{-1}\scovinv{\uds}{}$,
where the last bound is by Lemma~\ref{lem:sub-contam-bounded-cov}.
This implies $\norm{\Anorm\scov{\cds}{G \cap S}\Anorm} \leq \frac{\aop}{1-\aop}$.
Plugging this into \eqref{eq:removed-good-ub}, we have that
$\E[|G \cap (\Rnorm \setminus \Rend)|] \leq \frac{\aop}{1-\aop} \cdot \frac{n}{\wmax}$.
Because we assume $\tinner \geq \frac{8e\aop\log(\tnorm)}{1-\aop} \cdot \frac{1}{\eps}$,
the desired claim follows by combining this with \eqref{eq:total-removed-lb}.

\subsection{Proof of Lemma~\ref{lem:bounded-bad-covariance-main}}
\label{sec:pf-bounded-bad-cov}
\begin{lemma}[Anti-concentration]\label{lem:tail-lb}
  Let $X$ be a non-negative random variable with $\E[X] = \mu$.
  Then for all $\gamma > 0$, there exists $t \geq \mu/2$ such that
  \begin{align*}
    \P(X \geq t) \geq \frac{\gamma \mu}{2} t^{-(1+\gamma)}.
  \end{align*}
\end{lemma}
\begin{proof}
  Suppose for contradiction that for some $\gamma > 0$ and all $t \geq \mu/2$
  that $\P(X \geq t) < \frac{\gamma \mu}{2} t^{-(1+\gamma)}$.
  Then
  \begin{align*}
    \mu = \E[X] = \int_0^\infty \P(X \geq t) dt
    < \frac{\mu}{2} + \int_{\mu/2}^\infty \frac{\gamma \mu}{2} t^{-(1+\gamma)} dt
     & \Longrightarrow \frac{1}{\gamma} < \int_{\mu/2}^\infty t^{-(1+\gamma)} dt.
  \end{align*}
  However, the right-hand side is less than $1/\gamma$, yielding a contradiction.
\end{proof}

\begin{lemma}[Bounded number of negatively correlated points]
  \label{lemma:negative-correlation-bound}
  Let $v_1, \dots, v_k \in \R^p$ such that for all distinct $i,j \in [k]$,
  $\norm{v_i}^2 \leq M^2$ and $v_i^T v_j \leq -b$.
  Then $k \leq \frac{M^2}{b} + 1$.
\end{lemma}
\begin{proof}
  Observe that
  \begin{align*}
    0
     & \leq \bigg\|\sum_{i=1}^k v_i\bigg\|^2
    = \sum_{i=1}^k \norm{v_i}^2 + \sum_{i \neq j} v_i^T v_j
    \leq kM^2 - k(k-1)b.
  \end{align*}
  Rearranging yields the result.
\end{proof}

\begin{lemma}[Restatement of Lemma~\ref{lem:bounded-bad-covariance-main}]
  Let $\ell,\ell' \in \{1,2\}$ for $\ell \neq \ell'$. Let $m \geq \frac{12e \tnorm^2\log(\tnorm)}{\tinner}$. Then for all $u \in \R^d$ and $T \subseteq \Rnorm_\ell$ such that $|T| \leq m$ we have
  \begin{align*}
    u^T\scov{\cds}{T} u
    \leq
    \max\left\{
    \frac{4m}{n}\tinner \cdot u^T \scov{\cds}{\Rnorm_{\ell'}} u,
    16e \log(\tnorm)
    \cdot u^T \scov{\cds}{S_\ell} u
    \right\}.
  \end{align*}
\end{lemma}
\begin{proof} 
  The second claim of Lemma~\ref{lem:norm-filter-props} implies for all
  $i \in \Rnorm_\ell$ that
  $x_i \in \colspace(\scov{\cds}{\Rnorm_{\ell'}})$.
  Then because $T, S_\ell \subseteq \Rnorm_\ell$, we see that
  $\colspace(\scov{\cds}{T}),\colspace(\scov{\cds}{S_\ell})
    \subseteq \colspace(\scov{\cds}{\Rnorm_{\ell'}})$.
  This implies that it suffices to prove the claim for only
  $u \in \colspace(\scov{\cds}{\Rnorm_{\ell'}})$.

  Now, let $A = \scovisqrt{\cds}{\Rnorm_{\ell'}}$ for convenience.
  Rescaling $u \in \colspace(\scov{\cds}{\Rnorm_{\ell'}})$ by $A$,
  it suffices to show for all
  $v \in \S^{d-1} \cap \colspace(\scov{\cds}{\Rnorm_{\ell'}})$ that
  \begin{align}
    v^T A \scov{\cds}{T} A v
    \leq
    \max\left\{
    \frac{4m}{n}\tinner,
    16e \log(\tnorm)
    \cdot v^T A \scov{\cds}{S_\ell} A v
    \right\}. \label{eqn:bounded-bad-covariance-wts}
  \end{align}

  Again for convenience, let $\beta = v^T A \scov{\cds}{T} A v$.
  Equation~\eqref{eqn:bounded-bad-covariance-wts} is immediate for
  $\beta \leq \frac{4m}{n}\tinner$,
  so suppose $\beta > \frac{4m}{n}\tinner$.
  Let $I \sim \uniform(T)$. Then
  $\E[x_Ix_I^T] = \frac{n}{|T|}\scov{\cds}{T}$,
  which implies
  \begin{align*}
    \E[(x_I \cdot Av)^2]
    = \frac{n}{|T|} \cdot v^TA\scov{\cds}{T} Av
    = \frac{n \beta}{|T|}.
  \end{align*}
  Because $T \subseteq \Rnorm$, from Lemma~\ref{lem:norm-filter-props} we have for all $i \in T$ that
  \begin{align}
    (x_i \cdot Av)^2 \leq x_i^T A^2 x_i \leq \tnorm^2. \label{eqn:bounded-bad-covariance-1}
  \end{align}
  Lemma~\ref{lem:tail-lb} implies for all $\gamma > 0$ there exists
  $t \geq \frac{n \beta}{2|T|}$ such that
  \begin{align*}
    \P((x_I \cdot Av)^2 \geq t)
    \geq \frac{\gamma n\beta}{2|T|}t^{-(1+\gamma)}
    \geq \tnorm^{-2\gamma}\frac{\gamma n\beta}{2 |T| t},
  \end{align*}
  where the second inequality follows because
  \eqref{eqn:bounded-bad-covariance-1} implies $t \leq \tnorm^2$.
  Picking $\gamma = \frac{1}{2}\log^{-1}(\tnorm)$ yields
  $\P((x_I \cdot Av)^2 \geq t) \geq \frac{\gamma n \beta}{2e t |T|}$,
  which implies
  \begin{align*}
    |\{i \in T \mid x_i \cdot Av \geq \sqrt{t}\}|
    \geq \frac{\gamma \beta n}{4et}
    ~~\textrm{or}~~
    |\{i \in T \mid -x_i \cdot Av \geq \sqrt{t}\}|
    \geq \frac{\gamma \beta n}{4et}.
  \end{align*}
  We will prove the claim assuming the first case, as the second case is symmetric.

  Let $T' = \{i \in T \mid x_i \cdot Av \geq \sqrt{t}\}$.
  Then we have that
  \begin{align}
    v^T A\scov{\cds}{S_\ell}A v
    = \frac{1}{n} \sum_{i \in S_\ell} (x_i \cdot Av)^2
    \geq \frac{1}{n} \sum_{i \in T' \cap S_\ell} (x_i \cdot Av)^2
    \geq \frac{t}{n} |T' \cap S_\ell|.
    \label{eq:covS-lb}
  \end{align}
  We proceed to lower bound $|T' \cap S_\ell|$ in terms of $\beta$.
  Notice that for all $i, j \in T'$, we have $(x_i \cdot Av)(x_j \cdot Av) \geq t$.
  For distinct $i, j \in T' \setminus S_\ell$, the fact that
  $S_\ell$ contains neither $i$ nor $j$ implies $|x_i^T A^2 x_j| \leq \tinner$.
  Now observe the following identity that holds for all $x, y \in \R^d$:
  \begin{align*}
    x^T A^2 y
    = (x\cdot Av)(y \cdot Av) + (Ax - (x\cdot Av)v)\cdot(Ay - (y\cdot Av)v).
  \end{align*}
  Defining $y_i := (Ax_i - (x_i\cdot Av)v)$ for $i \in [n]$, we have
  for all distinct $i, j \in T' \setminus S_\ell$ that
  \begin{align*}
    (x_i \cdot Av)(x_j \cdot Av) + y_i \cdot y_j
    = x_i^T A^2 x_j
    \leq |x_i^T A^2 x_j|
    \leq \tinner.
  \end{align*}
  Combining this with the fact that $(x_i \cdot Av)(x_j \cdot Av) \geq t$,
  we thus see that $-y_i \cdot y_j \geq t - \tinner$.
  Because we assumed $|T| \leq m$ and $\beta > \frac{4m}{n}\tinner$,
  we have that $t \geq \frac{n \beta}{2|T|} \geq 2\tinner$,
  and so $-y_i \cdot y_j \geq t/2$.
  Also, the norm-filter step guarantees from
  Lemma~\ref{lem:norm-filter-props} that gives that
  $\norm{y_i}^2 \leq \norm{A x_i}^2 \leq \tnorm^2$ for all $i \in T'$.
  Thus, applying
  Lemma~\ref{lemma:negative-correlation-bound} to
  $\{y_i \mid i \in T' \setminus S_\ell\}$ gives us
  $|T' \setminus S_\ell| \leq \frac{2\tnorm^2}{t} + 1 \leq \frac{3\tnorm^2}{t}$.
  This implies
  \begin{align*}
    |T' \cap S_\ell| = |T'| - |T' \setminus S_\ell|
    \geq \frac{1}{t}\left(\frac{\gamma \beta n}{4e} - 3\tnorm^2\right)
    \geq \frac{\gamma \beta n}{8et},
  \end{align*}
  where the last inequality is by $\beta > \frac{4m}{n}\tinner$ and
  $m \geq \frac{12e \tnorm^2\log(\tnorm)}{\tinner}$.
  Combining this with \eqref{eq:covS-lb}, we have
  $v^T A\scov{\cds}{S_\ell}A v \geq \frac{\gamma \beta}{8e}
    = \frac{1}{16e\log(\tnorm)}v^TA\scov{\cds}{T} A v$,
  which is precisely what we wanted to show in
  \eqref{eqn:bounded-bad-covariance-wts}.
\end{proof}
\subsection{Proof of Lemma~\ref{lemma:few-good-points-in-S}}
\label{sec:pf-good-pts-in-S}
Recall that in our notation, for an arbitrary dataset $\ds = \{y_1, \dots, y_m\}
  \subset \R^d$, we denote $\scov{\ds}{} = \frac{1}{m}\sum_{i=1}^m y_iy_i^T$
and for $T \subseteq [m]$, we denote
$\scov{\ds}{T} = \frac{1}{m}\sum_{i \in T} y_i y_i^T$.

\begin{lemma}\label{lem:bound-large-norms}
  Let $\ds = \{y_1, \dots, y_m\} \subset \R^d$ such that for all $i \in [m]$,
  $y_i^T \scovinv{\ds}{} y_i \leq M^2$.
  Also let $\gamma, b > 0$, $T \subseteq [m]$ and
  $\cW = \{w_1, \dots, w_k\} \subset \R^d$ be such that
  $\scov{\cW}{} \preceq \frac{\gamma m}{k}\scov{\ds}{T}$ and for all $j \in [k]$,
  $w_j^T \big(\scovinv{\ds}{T} - \scovinv{\ds}{}\big) w_j \geq b$.
  Then $k \leq \frac{\gamma M^2}{b}(m - |T|)$.
\end{lemma}
\begin{proof}

  Let $J \sim \uniform([k])$ so that we have
  $\E[w_J w_J^T] = \scov{\cW}{}$.
  Then because $w_J^T (\scovinv{\ds}{T} - \scovinv{\ds}{}) w_J \geq b$ almost surely,
  we have that 
  \begin{align*}
    b
    \leq \E[w_J^T (\scovinv{\ds}{T} - \scovinv{\ds}{}) w_J]
    = \tr((\scovinv{\ds}{T} - \scovinv{\ds}{})\scov{\cW}{}).
  \end{align*}
  Notice 
  $\colspace(\scov{\cW}{}) \subseteq \colspace(\scov{\ds}{T})
  \subseteq \colspace(\scov{\ds}{})$,
  so that $\scov{\cW}{} \preceq \frac{\gamma m}{k}\scov{\ds}{T}$
  implies
  \begin{align*}
    \tr((\scovinv{\ds}{T} - \scovinv{\ds}{})\scov{\cW}{})
    \leq \frac{\gamma m}{k}\tr((\scovinv{\ds}{T} - \scovinv{\ds}{})\scov{\ds}{T}).
  \end{align*}
  To bound the trace, we first have
  \begin{align*}
    \tr((\scovinv{\ds}{T} - \scovinv{\ds}{})\scov{\ds}{T})
    = \tr(I - \scovinv{\ds}{}\scov{\ds}{T})
    = \tr(\scovinv{\ds}{}(\scov{\ds}{} - \scov{\ds}{T})).
  \end{align*}
  Then because $\scov{\ds}{} - \scov{\ds}{T}
    = \frac{1}{m} \sum_{i \in [m] \setminus T} y_i y_i^T$,
  we have
  \begin{align*}
    \tr(\scovinv{\ds}{}(\scov{\ds}{} - \scov{\ds}{T}))
     & = \frac{1}{m} \sum_{i \in [m] \setminus T} y_i^T \scovinv{\ds}{} y_i
    \leq \frac{M^2(m-|T|)}{m}.
  \end{align*}
  Combining all of the above displays and rearranging proves the claim.
\end{proof}

\begin{lemma}[Restatement of Lemma~\ref{lemma:few-good-points-in-S}]
  Suppose $\tinner \geq 2\binner$, $\tnorm \geq \bnorm$, and $|G \cap \Rnorm| \geq n - \min\{\mop, \frac{n}{4}\}$. Then we have
  \begin{align*}
    |G \cap S|
    \leq 4\mbig + |B \cap \Rnorm|
    + \frac{72 \tnorm^2}{\tinner} (n-|G \cap \Rnorm|).
  \end{align*}
\end{lemma}
\begin{proof}
  Throughout this proof, we denote $\Rnorm$ as $R$ for ease of notation.

  Now $S$ is a union of $S_1$ and $S_2$,
  where $S_1 = \computeS(\cds_{R_1}, \scov{\cds}{R_2}, \tinner)$
  and $S_2 = \computeS(\cds_{R_2}, \scov{\cds}{R_1}, \tinner)$.
  We note that $S_1 \subseteq R_1$ and $S_2 \subseteq R_2$.
  For $S_1 \subseteq R_1$, this is because $S_1$ is constructed from pairs
  $i, j \in [n]$ for which the
  corresponding entries in $\cds_{R_1}$ have
  $\scov{\cds}{R_2}$-normalized inner product exceeding $\tinner$. 
  However, for $i \in R_1$, the $i$th entry of $\cX_{R_1}$ is $0$,
  and so has an inner product of $0$ with all other entries of $\cX_{R_1}$.
  A symmetric argument appleis for $S_2 \subseteq R_2$.

  Thus, we have that $\Hone$ and $\Htwo$ partition $[n]$,
  $R_1$ and $R_2$ partition $R$, and
  $S_1$ and $S_2$ partition $S$.
  It therefore suffices to show that
  \begin{align*}
    |G \cap S_1|
     & \leq 2\mbig + |B \cap R_1|
    + \frac{72\tnorm^2}{\tinner}(|\Htwo| - |G \cap R_2|),\mbox{ and} \\
    |G \cap S_2|
     & \leq 2\mbig + |B \cap R_2|
    + \frac{72\tnorm^2}{\tinner}(|\Hone| - |G \cap R_1|).
  \end{align*}
  We will only show the first inequality, as the argument for the second
  inequality is symmetric.

  Because $S_1 = \computeS(\cds_{R_1}, \scov{\cds}{R_2}, \tinner)$
  is constructed by adding pairs of indices $i, j \in R_1$,
  we can, for every $i \in S_1$, define $j(i) \in S_1$ to be the other index
  that $i$ was added to $S_1$ alongside.
  Because the condition in Line~\ref{compS:loop-cond} of $\computeS$ prevents
  a point from being added to $S_1$ more than once, we have that the choice of
  $j(i)$ is unique for each $i \in S_1$ and that
  $j(i) \neq j(i')$ for distinct $i, i' \in S_1$.

  To bound $|G \cap S_1|$, first observe that for each $i \in G \cap S_1$,
  either $j(i) \in B \cap S_1$ or $j(i) \in G \cap S_1$.
  By injectivity of $j(\cdot)$, there are at most $|B \cap S_1|$
  points $i \in G \cap S_1$ are such that $j(i) \in B \cap S_1$.
  Moreover, because $S_1 \subseteq R_1$, we have
  $|B \cap S_1| \leq |B \cap R_1|$.
  Therefore, all we need to show is
  \begin{align*}
    |\{i \in G \cap S_1 \mid j(i) \in G \cap S_1\}|
    \leq 2\mbig + \frac{72 \tnorm^2}{\tinner} (|\Htwo|-|G \cap R_2|).
  \end{align*}

  Now consider the set
  $Q = \{i \in G \cap S_1 \mid j(i) \in G \cap S_1,
    |x_i^T\scovinv{\uds}{\Htwo}x_{j(i)}| > \binner\}$.
  Because $\cZ$ is
  $(\mbig, \binner)$-inner product bounded,
  there exists a set $T$ with $|T| \geq n-|\Htwo|-\mbig$ such that
  $|z_i^T\scovinv{\uds}{\Htwo}z_{j}| \leq \binner$ for all distinct
  $i, j \in T$.
  By inclusion-exclusion, we have that $|Q \cap T| \geq |Q| - \mbig$.
  Thus, if $|Q| \geq 2\mbig + 1$, we have by Pigeonhole Principle that
  for some $i \in Q$, both
  $i \in T$ and $j(i) \in T$.
  This contradicts the definition of $Q$ and so we have
  $|Q| \leq 2\mbig$.
  Thus, it only remains to show
  \begin{align*}
    |\{i \in G \cap S_1 \mid j(i) \in G \cap S_1,
    |x_i^T\scovinv{\uds}{\Htwo}x_{j(i)}| \leq \binner\}|
    \leq \frac{72 \tnorm^2}{\tinner} (|\Htwo|-|G \cap R_2|).
  \end{align*}

  Let $i \in G \cap S_1$ such that $j=j(i) \in G \cap S_1$
  and $|x_i^T\scovinv{\uds}{\Htwo}x_{j(i)}| \leq \binner$.
  We may upper bound $|x_i^T \scovinv{\cds}{R_2} x_j|$ as
  \begin{align*}
    |x_i^T \scovinv{\cds}{R_2} x_j|
     & \leq \Big|x_i^T \scovinv{\uds}{\Htwo} x_j\Big|
    + \Big|x_i^T \Big(\scovinv{\cds}{G \cap R_2} - \scovinv{\uds}{\Htwo}\Big) x_j\Big|         \\
     & \qquad+ \Big|x_i^T \Big(\scovinv{\cds}{G \cap R_2} - \scovinv{\cds}{R_2}\Big) x_j\Big|.
  \end{align*}
  The first term is bounded by $\binner$ by assumption.
  Now notice that
  $\scov{\cds}{G \cap R_2} = \scov{\uds}{G \cap R_2} \preceq \scov{\uds}{\Htwo}$ and
  $\scov{\cds}{G \cap R_2} \preceq \scov{\cds}{R_2}$, and so by applying Cauchy-Schwarz,
  \begin{align*}
    |x_i^T \scovinv{\cds}{R_2} x_j|
     & \leq \binner
    + \sqrt{x_i^T \Big(\scovinv{\cds}{G \cap R_2} - \scovinv{\uds}{\Htwo}\Big) x_i}
    \sqrt{x_j^T \Big(\scovinv{\cds}{G \cap R_2} - \scovinv{\uds}{\Htwo}\Big) x_j}          \\
     & \qquad+ \sqrt{x_i^T \Big(\scovinv{\cds}{G \cap R_2} - \scovinv{\cds}{R_2}\Big) x_i}
    \sqrt{x_j^T \Big(\scovinv{\cds}{G \cap R_2} - \scovinv{\cds}{R_2}\Big) x_j}.
  \end{align*}

  Combining this with the facts that
  $|x_i^T \scovinv{\cds}{R_2} x_j| > \tinner$ and $\binner \leq \frac{1}{2}\tinner$,
  we must have that
  \begin{enumerate}
    \item $x_i^T \big(\scovinv{\cds}{G \cap R_2} - \scovinv{\uds}{\Htwo}\big) x_i
            \geq \frac{1}{4}\tinner$; or
    \item $x_i^T \big(\scovinv{\cds}{G \cap R_2} - \scovinv{\cds}{R_2}\big) x_i
            \geq \frac{1}{4}\tinner$; or
    \item $x_j^T \big(\scovinv{\cds}{G \cap R_2} - \scovinv{\uds}{\Htwo}\big) x_j
            \geq \frac{1}{4}\tinner$; or
    \item $x_j^T \big(\scovinv{\cds}{G \cap R_2} - \scovinv{\cds}{R_2}\big) x_j
            \geq \frac{1}{4}\tinner$.
  \end{enumerate}

  Now define the sets
  \begin{align*}
    \Sgone & \defeq \big\{i \in G \cap R_1
    \mid x_i^T \big(\scovinv{\cds}{G \cap R_2} - \scovinv{\uds}{\Htwo}\big)x_i
    \geq \tfrac{1}{4}\tinner\big\},
    \\
    \Sgtwo & \defeq \big\{i \in G \cap R_1
    \mid x_i^T\big(\scovinv{\cds}{G \cap R_2} - \scovinv{\cds}{R_2}\big) x_i
    \geq \tfrac{1}{4}\tinner\}.
  \end{align*}
  We have shown that for $i \in G \cap S_1$ such that $j(i) \in G \cap S_1$,
  either $i \in \Sgone \cup \Sgtwo$ or $j(i) \in \Sgone \cup \Sgtwo$.
  Because $j(\cdot)$ is injective, this allows to conclude that
  $|G \cap S_1| \leq 2|\Sgone \cup \Sgtwo| \leq 2|\Sgone| + 2|\Sgtwo|$.
  Before bounding $|\Sgone|$ and $|\Sgtwo|$, we first note that
  by assumption on $|G \cap R|$, we have
  \begin{align*}
    |\Htwo \setminus (G \cap R_2)|
    \leq |[n] \setminus (G \cap R)|
    \leq \min\{\mop, \tfrac{n}{4}\}.
  \end{align*}
  Then because $\cZ$ is $(\mop, \aop)$-operator norm resilient,
  we have by Lemma~\ref{lem:sub-contam-bounded-cov} that
  $\scov{\uds}{\Htwo} \preceq (1-\aop)^{-1} \scov{\uds}{G \cap R_2}$.

  Now to bound $|\Sgone|$, we will apply Lemma~\ref{lem:bound-large-norms} to
  \begin{align*}
    \ds = \{z_i \mid i \in \Htwo\},
    ~~~~
    \cW = \{x_i \mid i \in \Sgone\},
    ~~~~\mbox{and}~~~~
    T = G \cap R_2.
  \end{align*}
  To do so, we first need to find the values of $M, C, b$ with
  which we can apply the lemma.
  Because $\cZ$ is $\bnorm$-norm bounded, we have
  for all $i \in \Htwo$ that
  \begin{align*}
    z_i^T \scovinv{\ds}{} z_i
    = \frac{|\Htwo|}{n} z_i^T \scovinv{\uds}{\Htwo} z_i
    \leq \frac{1}{2}\bnorm^2,
  \end{align*}
  so we may take $M = \frac{1}{\sqrt{2}}\bnorm$.
  By definition of $\Sgone$, we have for $i \in \Sgone$ that
  \begin{align*}
    x_i^T\big(\scovinv{\ds}{T} - \scovinv{\ds}{})x_i
    = \frac{|\Htwo|}{n} x_i^T\big(\scovinv{\uds}{G \cap R_2} - \scovinv{\uds}{\Htwo})x_i
    \geq \frac{1}{8}\tinner,
  \end{align*}
  so we may take $b = \frac{1}{8}\tinner$.
  Finally, to find $\gamma$, we first note that because $\cZ$ is $(\mop, \aop)$-operator
  norm resilient, we have that
  $\scov{\uds}{\Hone} \preceq \frac{3}{4}\scov{\uds}{}$.
  Since $\scov{\uds}{} = \scov{\uds}{\Hone} + \scov{\uds}{\Htwo}$, this implies that
  $\scov{\uds}{\Hone} \preceq 3\scov{\uds}{\Htwo}$.
  Thus, because $\Sgone \subseteq \Hone$, we have
  \begin{align*}
    \scov{\cW}{} = \frac{n}{|\Sgone|} \scov{\uds}{\Sgone}
    \preceq \frac{n}{|\Sgone|} \scov{\uds}{\Hone}
    \preceq \frac{3n}{|\Sgone|} \scov{\uds}{\Htwo}
    \preceq \frac{3n}{(1-\aop)|\Sgone|} \scov{\uds}{G \cap R_2}
    = \frac{3|\Htwo|}{(1-\aop)|\Sgone|} \scov{\ds}{T},
  \end{align*}
  and so we may take $\gamma = 3(1-\aop)^{-1}$.
  Applying Lemma~\ref{lem:bound-large-norms} with these quantities, we can
  conclude that
  $|\Sgone|
    \leq \frac{3}{1-\aop} \cdot \frac{\bnorm^2}{2} \cdot \frac{8}{\tinner} \cdot (|\Htwo| - |G \cap R_2|)
    \leq \frac{12\tnorm^2}{(1-\aop)\tinner}(|\Htwo| - |G \cap R_2|)$.

  To bound $|\Sgtwo|$, we will similarly apply
  Lemma~\ref{lem:bound-large-norms} to
  \begin{align*}
    \ds = \{x_i \mid i \in R_2\},
    ~~~~
    \cW = \{x_i \mid i \in \Sgtwo\},
    ~~~~\mbox{and}~~~~
    T   = G \cap R_2.
  \end{align*}
  Again, we need to find the values of $M, C, b$ with which we can
  apply Lemma~\ref{lem:bound-large-norms} to these datasets.
  For $M$, we have by Lemma~\ref{lem:norm-filter-props} that for all $i \in R_2$,
  \begin{align*}
    x_i^T \scovinv{\ds}{} x_i
    = \frac{|R_2|}{n} x_i^T \scovinv{\cds}{R_2} x_i
    \leq \frac{1}{2}\tnorm^2,
  \end{align*}
  so we may take $M = \frac{1}{\sqrt{2}}\tnorm$.
  By definition of $\Sgtwo$, we have for $i \in \Sgtwo$ that
  \begin{align*}
    x_i^T\big(\scovinv{\ds}{T} - \scovinv{\ds}{})x_i
    = \frac{|R_2|}{n} x_i^T\big(\scovinv{\cds}{G \cap R_2} - \scovinv{\cds}{R_2})x_i
    \geq \frac{|R_2|}{4n}\tinner.
  \end{align*}
  Now note that
  \begin{align*}
    |R_2|
    \geq |G \cap R_2|
    = |\Htwo| - |\Htwo \setminus (G \cap R_2)|
    \geq \frac{n}{2} - \min\left\{\mop, \frac{n}{4}\right\}
    \geq \frac{n}{4},
  \end{align*}
  so we may take $b = \frac{1}{16}\tinner$.
  Finally, to find $\gamma$, by the same argument as when we computed
  $\gamma$ for $|\Sgone|$, we have
  that
  \begin{align*}
    \scov{\cW}{} \preceq \frac{3n}{(1-\aop)|\Sgtwo|} \scov{\uds}{G \cap R_2}
    =  \frac{3n}{(1-\aop)|\Sgtwo|} \scov{\cds}{G \cap R_2}
    = \frac{3|R_2|}{(1-\aop)|\Sgtwo|}\scov{\ds}{T},
  \end{align*}
  and so we may take $\gamma = 3(1-\aop)^{-1}$.
  Applying Lemma~\ref{lem:bound-large-norms} with these quantities, we can
  conclude that
  $|\Sgtwo|
    \leq \frac{3}{1-\aop} \cdot \frac{\tnorm^2}{2} \cdot \frac{16}{\tinner}
    \cdot (|R_2| - |G \cap R_2|)
    \leq \frac{24\tnorm^2}{\tinner}(|\Htwo| - |G \cap R_2|)$.
  Putting it all together, we have
  $2|\Sgone| + 2|\Sgtwo| \leq \frac{72\tnorm^2}{(1-\aop)\tinner}(\Htwo - |G \cap R_2|)$,
  proving the claim.
\end{proof}

\section{Proofs for Section~\ref{sec:subg-perf}}\label{sec:subg-perf-pfs}

\subsection{Proof of Lemma~\ref{lem:subg-goodness}}\label{sec:lem-subg-goodness-pf}



We split the proof into two lemmas.
Lemma~\ref{lem:subg-norm-ip-bddness} shows that $\bnorm$-norm and
$(0, \binner)$-inner product boundedness
hold with probability at least $1-3\delta$.
Lemma~\ref{lem:subg-op-norm-resilience} shows that
$(\mop, \aop)$-operator norm resilience
with probability at
least $1-4\delta$.
A union bound then proves the claim.

\begin{lemma}[Norm and inner product boundedness]\label{lem:subg-norm-ip-bddness}
  Suppose $n \gtrsim \max\{\sigma^4 d + \log(1/\delta),
    \sigma^2(d+\log(n/\delta))\}$.
  Then there exists $\bnorm$ and $\binner$ satisfying
  \begin{align*}
    \bnorm^2 = O(\sigma^2 \cdot (d + \log(n/\delta)))
    ~~~~\mbox{and}~~~~
    \binner  = \sigma\sqrt{2\log(n^2/\delta)}\bnorm 
  \end{align*}
  such that
  with probability at least $1-3\delta$,
  $\uds$ is $\bnorm$-norm bounded and $(0, \binner)$-inner product bounded.
\end{lemma}
\begin{proof}
  Expanding out the definitions of norm and inner product boundedness,
  we need to show that there exists $\bnorm, \binner$ satisfying
  $\bnorm^2 = O(\sigma^2 \cdot (d + \log(n/\delta)))$ and
  $\binner  = \sigma\sqrt{2\log(n^2/\delta)}\bnorm$
  such that with probability at least $1-3\delta$,
  \begin{align*}
    \max_{\ell \in \{1, 2\}} \sup_{i \in [n]} Z_i^T \scovinv{\uds}{\Hell} Z_i
    \leq \bnorm^2
    ~~~~\mbox{and}~~~~
    \max_{\ell \in \{1, 2\}}
    \sup_{\substack{i,j \in [n] \setminus \Hell \\ i \neq j}}
    |Z_i^T \scovinv{\uds}{\Hell} Z_j| \leq \binner.
  \end{align*}
  We begin by using independence and subgaussianity to bound the
  inner products between distinct points in terms of the norms of the points.
  For $\ell \in \{1, 2\}$ and distinct $i,j \in [n] \setminus \Hell$,
  notice that $\uds_{-i}$ determines $\scovinv{\uds}{\Hell}Z_j$.
  Therefore, fixing $\uds_{-i}$, we have over the randomness of $Z_i$ that
  $Z_i^T\scovinv{\uds}{\Hell} Z_j$ is subgaussian with variance proxy
  $\sigma^2 \big\|\Sigma^{1/2}\scovinv{\uds}{\Hell} Z_j\big\|^2$.
  A subgaussian tail bound
  (Lemma~\ref{lem:subg-tail-bd}) gives that with
  probability at least $1-\beta$ over the randomness of $Z_i$,
  \begin{align*}
    |Z_i^T\scovinv{\uds}{\Hell} Z_j|
     & \leq \sigma\sqrt{\log(e/\beta)}
    \big\|\Sigma^{1/2}\scovinv{\uds}{\Hell} Z_j\big\| \\
     & \leq \sigma\sqrt{1 + \log(1/\beta)}
    \big\|\Sigma^{1/2}\scovisqrt{\uds}{\Hell}\big\|
    \|\scovisqrt{\uds}{\Hell} Z_j\|.
  \end{align*}
  Taking $\beta = \frac{\delta}{n^2}$ and a union bound,
  we have that the above holds simultaneously for all
  $\ell \in \{1, 2\}$ and
  all distinct pairs $i, j \in [n] \setminus \Hell$.

  Now suppose we further condition on the events that
  \begin{align*}
    \max_{\ell \in \{1, 2\}} \dpsd(\scov{\uds}{\Hell}, \Sigma) \leq \frac{1}{2}
  \end{align*}
  and
  \begin{align*}
    \sup_{i \in [n]} \norm{Z_i}_{\Sigma}^2
    \leq \sigma^2 \cdot O(d + \log(n/\delta))
    = \frac{1}{2} \bnorm^2,
  \end{align*}
  both of which happen with probability at least $1-\delta$ by
  Lemmas~\ref{lem:subg-emp-cov-conc} and~\ref{lem:subg-norm-boundedness},
  respectively.
  We will now show that both of the desired claims hold.

  First note that $\dpsd(\scov{\uds}{\Hell}, \Sigma) \leq \frac{1}{2}$
  implies $\scov{\uds}{\Hell} \succeq \frac{1}{2}\Sigma$.
  This then implies both
  \begin{align*}
    \big\|\Sigma^{1/2}\scovinv{\uds}{\Hell}\Sigma^{1/2}\big\| \leq 2
    ~~~\mbox{and}~~~
    \|\Sigma^{1/2}\scovisqrt{\uds}{\Hell}\| \leq \sqrt{2}.
  \end{align*}

  Then the first claim follows immediately by Cauchy-Schwarz because
  \begin{align*}
    \max_{\ell \in \{1, 2\}} \sup_{i \in [n]} Z_i^T\scovinv{\uds}{\Hell} Z_i
    \leq \max_{\ell \in \{1, 2\}} \sup_{i \in [n]}
    \big\|\Sigma^{1/2}\scovinv{\uds}{\Hell}\Sigma^{1/2}\big\|
    \norm{\Sigma^{\dag/2} Z_i}^2
    \leq \bnorm^2.
  \end{align*}

  The second claim also follows immediately because
  for all $\ell \in \{1, 2\}$ and distinct pairs $i, j \in [n] \setminus \Hell$,
  our earlier bound gives
  \begin{align*}
    |Z_i^T\scovinv{\uds}{\Hell} Z_j|
     & \leq \sigma\sqrt{1 + \log(n^2/\delta)}
    \big\|\Sigma^{1/2}\scovisqrt{\uds}{\Hell}\big\|
    \|\scovisqrt{\uds}{\Hell} Z_j\|                     \\
     & \leq \sigma\sqrt{2(1 + \log(n^2/\delta))}\bnorm.
  \end{align*}

  This proves the claim.
\end{proof}

We adapt the following argument for operator norm resilience of subgaussian data
from \citet[Proposition 3.3]{DiakonikolasKa22}.
\begin{lemma}[Subgaussian operator norm resilience]\label{lem:subg-op-norm-resilience}
  Let $n \gtrsim \sigma^4 d + \log(1/\delta)$ and
  $\mop \leq n$.
  Then there exists $\aop$ satisfying
  \begin{align*}
    \aop
    = 2\left(1 + 2\sigma^2\log\left(\frac{n}{2\mop}\right)\right) \cdot \frac{\mop}{n}
    + O\left(\sigma^2 \sqrt{\frac{d + \log(1/\delta)}{n}}\right).
  \end{align*}
  such that
  with probability at least $1-2\delta$,
  $\cZ$ is $(\mop, \aop)$-operator norm resilient.
\end{lemma}
\begin{proof}
  Let $H \in \{\Hone, \Htwo\}$ and for ease of notation,
  let $m = |H| = \frac{n}{2}$.
  We will show the required properties for $H$ hold with probability
  at least $1-2\delta$, so that the claim follows by a union bound over $H$.

  We first show that with probability at least $1-\delta$,
  $\scov{\uds}{\genset} \preceq \frac{3}{4}\scov{\uds}{}$.
  Because $n \geq m \gtrsim \sigma^4d + \log(4/\delta)$,
  we have by Lemma~\ref{lem:subg-emp-cov-conc} that probability $1-\delta$,
  \begin{align*}
    \dpsd\left(\frac{1}{m}\sum_{i \in \genset} Z_iZ_i^T, \Sigma\right) \leq c
    ~~~\mbox{and}~~~
    \dpsd\left(\frac{1}{n}\sum_{i=1}^n Z_iZ_i^T, \Sigma\right)
    \leq c,
  \end{align*}
  where $c > 0$ is a small constant we will set later.
  This implies $\frac{1}{m}\sum_{i \in \genset} Z_iZ_i^T \preceq (1+c) \Sigma$
  and $\frac{1}{n}\sum_{i=1}^n Z_iZ_i^T \succeq (1-c)\Sigma$ so that
  \begin{align*}
    \frac{1}{m}\sum_{i \in S} Z_iZ_i^T \preceq \frac{(1+c)}{(1-c)n}\sum_{i=1}^n Z_iZ_i^T
    \Longrightarrow 2\scov{\uds}{\genset} \preceq \frac{1+c}{1-c}\scov{\uds}{}.
  \end{align*}
  Picking $c = \frac{1}{5}$, we obtain
  $\scov{\uds}{\genset} \preceq \frac{3}{4}\scov{\uds}{}$,
  proving the first requirement.

  Now we show that with probability at least $1-\delta$,
  it holds that
  \begin{align*}
    \frac{1}{n}\sum_{i \in T} Z_i Z_i^T \preceq \aop \scov{\uds}{\genset}
  \end{align*}
  for all $T \subseteq \genset$ such that $|T| \leq \mop$.

  By
  $\dpsd(\frac{1}{m}\sum_{i \in \genset} Z_i Z_i^T, \Sigma) \leq \frac{1}{5}$,
  we know that
  $\frac{1}{m}\sum_{i \in \genset} Z_i Z_i^T \succeq \frac{1}{2} \Sigma$.
  Noting that $\scov{\uds}{\genset} = \frac{1}{2m}\sum_{i \in \genset} Z_i Z_i^T$,
  it thus suffices to show that with probability at least $1 - \delta$,
  \begin{align*}
    \frac{1}{m}\sum_{i \in T} Z_i Z_i^T \preceq \frac{\aop}{2} \Sigma
  \end{align*}
  for all $T \subseteq \genset$ such that $|T| \leq \mop$.
  Equivalently, we want to show that with
  probability at least $1-\delta$, it holds for all
  $v \in \S^{d-1} \cap \colspace(\Sigma)$ that
  \begin{align*}
    \sup_{\substack{T \subseteq \genset \\|T|=\mop}}
      \frac{1}{m} \sum_{i \in T} (v^T \Sigma^{\dag/2}Z_i)^2
    = \sup_{\substack{T \subseteq \genset \\|T|=\mop}}
      v^T \Sigma^{\dag/2}\left[\frac{1}{m} \sum_{i \in T} Z_i Z_i^T\right]
        \Sigma^{\dag/2} v
    \leq \frac{\aop}{2}.
  \end{align*}

  Let $\cV$ be a $\frac{1}{4}$-net of $\S^{d-1} \cap \colspace(\Sigma)$,
  which we can take such
  that $|\cV| \leq 9^d$ by \citet[Lemma 5.2]{Vershynin12}.
  Then \citet[Lemma 5.4]{Vershynin12} gives that
  \begin{align*}
    \sup_{v \in \S^{d-1}}
    \sup_{\substack{T \subseteq \genset \\|T|=\mop}} \frac{1}{m}
    \sum_{i \in T} (v^T \Sigma^{\dag/2} Z_i)^2
    \leq
    2\sup_{v \in \cV}
    \sup_{\substack{T \subseteq \genset \\|T|=\mop}} \frac{1}{m}
    \sum_{i \in T} (v^T \Sigma^{\dag/2} Z_i)^2.
  \end{align*}
  For each $v \in \cV$, we will show that
  \begin{align*}
    \sup_{\substack{T \subseteq \genset \\|T|=\mop}} \frac{1}{m}
    \sum_{i \in T} (v^T \Sigma^{\dag/2} Z_i)^2
    \leq (1 + 2\sigma^2\log(m/\mop)) \cdot \frac{\mop}{m}
    + O\left(\sigma^2 \sqrt{\frac{d + \log(1/\delta)}{m}}\right)
  \end{align*}
  with probability at least $1-9^{-d}\delta$,
  so that the claim follows by a union bound over all $v \in \cV$.

  For a fixed $v \in \cV$, let $Y_i = (v^T \Sigma^{\dag/2} Z_i)^2$.
  We then have that $Y_1, \dots, Y_n$ are i.i.d.\ $\sigma^2$-sub-exponential
  random variables.
  Let $M > 0$ be a threshold we specify later.
  Then,
  \begin{align*}
    \sup_{\substack{T \subseteq \genset                                          \\ |T|=\mop}} \frac{1}{m} \sum_{i \in T} Y_i
     & \leq \sup_{\substack{T \subseteq \genset                                  \\ |T|=\mop}}
    \frac{1}{m} \sum_{i \in T} Y_i \1{Y_i \leq M}
    + \frac{1}{m} \sum_{i \in \genset} Y_i \1{Y_i > M}                           \\
     & \leq \frac{\mop}{m} \cdot M + \frac{1}{n} \sum_{i \in S} Y_i \1{Y_i > M}.
  \end{align*}
  We now bound $\frac{1}{m} \sum_{i \in \genset} Y_i \1{Y_i > M}$ with high probability.
  Let $\mu = \E[Y_i \1{Y_i > M}]$.
  Because $Y_i$ is $\sigma^2$-sub-exponential, we first have that
  \begin{align*}
    \mu
     & = \E[Y_i \1{Y_i > M}]                                                \\
     & = \int_0^\infty \P(Y_i \1{Y_i > M} > t) dt                           \\
     & = \int_0^\infty \P(Y_i > \max\{t, M\}) dt                            \\
     & \leq \int_0^\infty \exp(-\max\{t, M\}/\sigma^2) dt.                  \\
     & = \int_0^M \exp(-M/\sigma^2) dt + \int_M^\infty \exp(-t/\sigma^2) dt \\
     & = (M+1)\exp(-M/\sigma^2).
  \end{align*}
  Then because $Y_i \1{Y_i > M}$ is also $\sigma^2$-subexponential, we have by
  a Bernstein-type inequality (Lemma~\ref{lem:bernstein}) that
  \begin{align*}
    \P\left(\frac{1}{n} \sum_{i \in \genset} [Y_i \1{Y_i > M}- \mu] \geq t\right)
     & = \P\left(\sum_{i \in \genset} [Y_i \1{Y_i > M} - \mu] \geq t n\right)                      \\
     & \leq \exp\left(-c \min\left\{\frac{(m t)^2}{\sigma^4m}, \frac{mt}{\sigma^2}\right\}\right).
  \end{align*}
  Picking
  \begin{align*}
    t
    \gtrsim \sigma^2 \left[\sqrt{\frac{d + \log(1/\delta)}{m}}
      + \frac{d + \log(1/\delta)}{m}\right]
    \asymp \sigma^2 \sqrt{\frac{d + \log(1/\delta)}{m}},
  \end{align*}
  we have
  $\P\left(\frac{1}{m} \sum_{i \in \genset} [Y_i \1{Y_i > M}- \mu] \geq t\right)
    \leq 9^{-d}\delta$.
  Thus, with probability at least $1 - 9^{-d}\delta$,
  \begin{align*}
    \sup_{\substack{T \subseteq \genset                    \\|T|=\mop}} \frac{1}{m} \sum_{i \in T} Y_i
     & \leq \frac{\mop}{m}\cdot M + (M+1)\exp(-M/\sigma^2)
    + O\left(\sigma^2 \sqrt{\frac{d + \log(1/\delta)}{m}}\right).
  \end{align*}
  By picking $M = \sigma^2 \log(m/\mop)$,
  we prove the claim.
\end{proof}


\subsection{Proof of Corollary~\ref{cor:subg-alg-perf}}\label{sec:subg-alg-perf-pf} 
For $\mbig = 0$ and any $\mop \leq n$,
Lemma~\ref{lem:subg-goodness} gives that
there exists $\bnorm, \binner, \aop$ satisfying
\begin{align*}
  \bnorm^2 & = \sigma^2 \cdot O(d + \log(n/\delta)) \\
  \binner  & = 2\sigma\sqrt{2\log(n^2/\delta)}\bnorm                       \\
  \aop     & = 2(1 + 2\sigma^2\log(n/2\mop)) \cdot \frac{\mop}{n}
  + O\left(\sigma^2 \sqrt{\frac{d + \log(1/\delta)}{n}}\right)
\end{align*}
such that
with probability at least $1-7\delta$,
the uncontaminated dataset
$\uds$ is $(\bnorm, \mbig, \binner, \mop, \aop)$-good.
To apply Theorem~\ref{thm:alg-result} with $C > 0$,
we need $\tinner, \tnorm$ to satisfy
\ExecuteMetaData[\SectionsPath subg-main-result.tex]{alg-reqs}
Picking $\tnorm = (1-\aop)^{-\frac{1}{2}}\bnorm$, we then need to pick
$\tinner$ satisfying
\begin{align}
  \tinner
  \gtrsim \max\left\{\binner,
  \frac{\aop\log(\tnorm)}{(1-\aop)\eps},
  \frac{\tnorm^2\log(\tnorm)}{\eps n},
  \frac{\tnorm^2 \eps n}{\mop}
  \right\}.\label{eq:subg-tinner-lb-1}
\end{align}

Because $\log(n/\delta) \lesssim d$ and $n \gtrsim \sigma^4 d$,
we have that for a sufficiently small
constant $c > 0$
and $\mop \leq \frac{cn}{\sigma^2\log \sigma^2}$,
the choice of $\aop$ satisfies
\begin{align*}
  \aop
  = \sigma^2 \cdot
    O\left(\frac{\mop}{n}\log\left(\frac{n}{\mop}\right) + \sqrt{\frac{d}{n}}\right)
  \leq \frac{1}{2}.
\end{align*}
This then implies $\tnorm \asymp \sigma \sqrt{d}$ and we can simplify \eqref{eq:subg-tinner-lb-1}
\begin{align*}
  \tinner
  \gtrsim \max\Big\{\sigma^2 \sqrt{d\log(n/\delta)},
  \frac{\aop \log(\sigma^2 d)}{\eps},
  \frac{\sigma^2 d\log(\sigma^2 d)}{\eps n},
  \frac{\sigma^2 \eps d n}{\mop}
  \Big\}.
\end{align*}
Notice that we require $\tinner \gtrsim \frac{\aop}{\eps}$,
so that Theorem~\ref{thm:alg-result}
gives an error bound of
\begin{align*}
  \err(\SigmaAlg, \scov{\uds}{})
   & \lesssim \aop + \eps \tinner
  \lesssim \eps \tinner.
\end{align*}
with probability at least $1-C$.

Because $\eps
  \lesssim \frac{\log n}{(\sigma^2\log \sigma^2)\sqrt{d}}$,
we can pick $\mop \asymp \frac{\eps\sqrt{d}}{\log n} n$
satisfying $\mop \leq \frac{cn}{\sigma^2\log \sigma^2}$.
This yields
\begin{align*}
  \aop
  \lesssim \sigma^2\sqrt{d}\left(\eps
    + \sqrt{1/n}\right)
\end{align*}
and allows us to choose
\begin{align*}
  \tinner
  &\asymp
  \sigma^2\max\left\{
    \sqrt{d\log(n/\delta)},
    \left(1
    + \frac{1}{\eps\sqrt{n}}\right)\sqrt{d}\log n,
    \frac{d\log n}{\eps n},
    \sqrt{d}\log n
  \right\} \\
  &\asymp
  \sigma^2\log(n)\sqrt{d} \max\left\{1, \frac{1}{\eps \sqrt{n}}\right\}.
\end{align*}
This then gives the error
\begin{align*}
  \err(\SigmaAlg, \scov{\uds}{})
  &\lesssim \eps \tinner
  \lesssim \sigma^2\log(n) \cdot (\eps\sqrt{d} + \sqrt{d/n}),
\end{align*}
proving the claim.
\section{Proofs for Section~\ref{sec:heavy-perf}}\label{sec:heavy-perf-pfs}
\subsection{Useful properties of \texorpdfstring{$\bar{P}$}{barP}}\label{sec:barP-props}
Here we establish some properties of $\bar{P}$ that will be useful in
proofs for moment bounded data.
By Lemma~\ref{lem:clipped-cov-err}, we know that
\begin{align*}
  \dpsd(\bar{\Sigma}, \Sigma)
  \leq m_k^k \left(\frac{\sqrt{d}}{\bclip}\right)^{k-2}.
\end{align*}
Then, if $\bclip \gtrsim m_k^{\frac{k}{k-2}}\sqrt{d}$,
we have that $\dpsd(\bar{\Sigma}, \Sigma) \leq \frac{1}{3}$.
Because $\norm{\bar{Z}}_\Sigma \leq \bclip$ almost surely,
we have by Lemma~\ref{lem:dpsd-prop} that
\begin{align}
  \norm{\bar{Z}}_{\bar{\Sigma}}
  \leq \norm{\bar{\Sigma}^{\dag/2}\Sigma^{1/2}}\norm{\bar{Z}}_\Sigma
  \leq \sqrt{\frac{3}{2}} \bclip,
  \label{eq:Pbar-as-bd}
\end{align}
almost surely.
Also, by moment-boundedness, we have for all $v \in \S^{d-1}$ that
\begin{align}
  \E[|\bar{\Sigma}^{\dag/2}\bar{Z} \cdot v|^k]^{\frac{1}{k}}
  \leq \E[|\bar{\Sigma}^{\dag/2}Z \cdot v|^k]^{\frac{1}{k}}
  \leq m_k(v^T\bar{\Sigma}^{\dag/2}\Sigma \bar{\Sigma}^{\dag/2} v)^{1/2}
  \leq \sqrt{\frac{3}{2}} m_k.
  \label{eq:Pbar-moment-bd}
\end{align}

\subsection{Proof of Lemma~\ref{lem:preproc-prop}}
\label{sec:preproc-prop}
We begin with a deterministic lemma about $\preprocess$.
Showing the preconditions of the lemma hold with high probability, we will
have proven Lemma~\ref{lem:preproc-prop}.
Before stating and proving that lemma, we first recall the following
lemma that we used
to prove Lemma~\ref{lemma:few-good-points-in-S}.
\begin{lemma}[Restatement of Lemma~\ref{lem:bound-large-norms}]
  \label{lem:bound-large-norms-heavy}
  Let $\cY = \{y_1, \dots, y_m\} \subset \R^d$ such that for all $i \in [m]$,
  $y_i^T \scovinv{\ds}{} y_i \leq M^2$.
  Also let $\gamma, b > 0$, $T \subseteq [m]$ and
  $\cW = \{w_1, \dots, w_k\} \subset \R^d$ be such that
  $\scov{\cW}{} \preceq \frac{\gamma m}{k}\scov{\ds}{T}$ and for all $j \in [k]$,
  $w_j^T \big(\scovinv{\ds}{T} - \scovinv{\ds}{}\big) w_j \geq b$.
  Then $k \leq \frac{\gamma M^2}{b}(m - |T|)$.
\end{lemma}

\begin{lemma}\label{lem:preproc-deterministic}
  Let $\Sigma$ be a PSD matrix and
  let $\tclip, \bclip > 0$ such that $\tclip \leq \frac{1}{\sqrt{8}}\bclip$.
  Let $\uds = \{z_1,\dots,z_n\}$ such that
  \begin{align*}
    \inf\{z_i^T\scovinv{\uds}{}z_i
      \mid i \in [n], \norm{z_i}_\Sigma > \bclip\}
    > \frac{1}{4}\bclip^2.
  \end{align*}
  Let $\bar{\uds} = \{\bar{z}_1, \dots, \bar{z}_n\}$,
  where $\bar{z}_i = z_i\1{\norm{z_i}_\Sigma \leq \bclip}$, and
  suppose that $\dpsd(\scov{\bar{\uds}}{}, \Sigma) \leq \frac{1}{2}$.
  Let $m = |\{i \in [n] \mid \norm{z_i}_\Sigma \geq \tclip/\sqrt{8}\}|$
  and suppose that $\bar{\uds}$ is
  $(m + \eps n, \frac{1}{2})$-operator norm resilient.
  Let $\cX = \{x_1,\dots,x_n\}$ be an $\eps$-contamination of
  $\uds$ and let $\cY = \preprocess(\cX,\tclip)$.
  Then there exists $T \subseteq [n]$ with $|T| \leq m$
  such that $\cY$ is a
  $\Big(1 + \frac{8n}{\bclip^2}\Big)\eps$-contamination of $\bar{\uds}_{[n]\setminus T}$.
\end{lemma}
\begin{proof}
  Let $G = \{i \in [n] \mid z_i = x_i\}$ and
  $\overline{G} = \{i \in [n] \mid \bar{z}_i = x_i\}$.
  Note that $|G| \geq (1-\eps)n$
  because $\cX$ is an $\eps$-contamination of $\uds$.
  Moreover, because $\uds$ and $\bar{\uds}$ only differ on points where
  $\norm{z_i}_\Sigma \geq \bclip \geq \tnorm/\sqrt{2}$, we also have that
  $|\bar{G}| \geq n - (m + \eps n)$.

  Now define the set
  \begin{align*}
    T = \{i \in [n] \mid z_i^T \scovinv{\cds}{}z_i \geq \tclip^2 \mbox{ or}
    \norm{z_i}_\Sigma > \bclip\}.
  \end{align*}
  We first show that $|T| \leq m$.
  For $i \in [n]$ such that $z_i^T \scovinv{\cds}{}z_i \geq \tclip^2$,
  \begin{align*}
    \tclip^2
    \leq z_i^T \scovinv{\cds}{}z_i
    \leq z_i^T \scovinv{\cds}{\Hone \cap \overline{G}}z_i
    = z_i^T \scovinv{\bar{\uds}}{\Hone \cap \overline{G}}z_i
    \leq 2 z_i^T \scovinv{\bar{\uds}}{\Hone}z_i,
  \end{align*}
  where the last inequality is by $(m+\eps n, \frac{1}{2})$-operator norm
  resilience of $\bar{\uds}$.
  Operator norm resilience and also gives that $z_i^T \scovinv{\bar{\uds}}{\Hone}z_i \leq 2 z_i^T \scovinv{\bar{\uds}}{}z_i \leq
  4 z_i^T \Sigma^\dag z_i$, which altogether imply that $\norm{z_i}_\Sigma \geq \tclip/\sqrt{8}$.
  Because we also assume that $\bclip \geq \sqrt{8}\tclip$,
  we have that $T \subseteq \{i \in [n] \mid \norm{z_i}_\Sigma \geq \tclip/\sqrt{2}\}$,
  which implies that $|T| \leq m$.

  Thus, all we need to show now is that $\cY = \{y_1, \dots, y_n\}$ is a
  $\Big(\frac{9n\eps}{\bclip^2}\Big)$-contamination of $\bar{\uds}_{[n]\setminus T}$.
  Note for any $i \in [n]$,
  \begin{align*}
    y_i
     & = x_i \1{x_i^T \scovinv{\cds}{} x_i \leq \tclip^2}           \\
    \bar{z_i}\1{i \not\in T}
     & = z_i \1{z_i^T \scovinv{\cds}{}z_i \leq \tclip^2 \mbox{ and}
      \norm{z_i}_\Sigma \leq \bclip},
  \end{align*}
  where the first line is by construction of $\preprocess$.
  Then for any $i \in [n]$ such that $y_i \neq \bar{z_i}\1{i \not\in T}$,
  we must have that
  \begin{enumerate}
    \item $x_i \neq z_i$ (in which case $i \not\in G$), or
    \item $x_i = z_i$ and $z_i^T \scovinv{\cds}{}z_i \leq \tclip^2$
          and $\norm{z_i}_\Sigma > \bclip$.
  \end{enumerate}
  The first case only occurs for at most $\eps n$ indices $i \in [n]$, and so
  letting $S$ denote the set of indices for which the second case holds,
  we turn to bounding $|S|$.

  First note that $S \subseteq G$, so that
  $\scov{\uds}{S} = \scov{\cds}{S} \preceq \scov{\cds}{G}$.
  Now, we have by assumption that for all $i \in [n]$ satisfying
  $\norm{z_i}_\Sigma > \bclip$, it holds that
  \begin{align*}
    \frac{1}{4}\bclip^2
    < z_i^T\scovinv{\uds}{}z_i
    \leq z_i^T\scovinv{\uds}{G}z_i
    = z_i^T\scovinv{\cds}{G}z_i.
  \end{align*}
  Because we also assume that $\tclip \leq \frac{1}{\sqrt{8}}\bclip$,
  we therefore have for $i \in S$ that
  \begin{align*}
    z_i^T \scovinv{\cds}{}z_i
    \leq \tclip^2
    \leq \frac{1}{8}\bclip^2
    \leq \frac{1}{2} z_i^T\scovinv{\cds}{G}z_i,
  \end{align*}
  and so
  \begin{align*}
    z_i^T\big(\scovinv{\cds}{G}-\scovinv{\cds}{}\big)z_i
    \geq \frac{1}{2}z_i^T\scovinv{\cds}{G}z_i
    \geq \frac{1}{8}\bclip^2.
  \end{align*}
  We now wish to apply Lemma~\ref{lem:bound-large-norms-heavy}
  with $\cY = \cX$, $\cW = \{z_i \mid i \in S\}$,
  and $T = G$.
  The above argument implies that we can take $b = \frac{1}{8}\bclip^2$.
  We can also take $M^2 = n$ because for all $i \in [n]$,
  $x_i^T\scovinv{\cds}{}x_i \leq x_i^T\left(\frac{1}{n} x_ix_i^T\right)^{\dag}x_i
    \leq n$.
  Finally we can also take $\gamma = 1$ because
  \begin{align*}
    \scov{\cW}{}
    = \frac{n}{|S|}\scov{\uds}{S}
    \preceq \frac{n}{|S|}\scov{\cds}{G}.
  \end{align*}
  Thus, applying Lemma~\ref{lem:bound-large-norms-heavy}, we conclude that
  $|S| \leq \frac{8n}{\bclip^2} (n - |G|) \leq \Big(\frac{8n}{\bclip^2}\Big)\eps n$.

  Putting it all together, we have proven that
  $\cY$ is a $\Big(1 + \frac{8n}{\bclip^2}\Big)\eps$-contamination of
  $\bar{\uds}_{[n]\setminus T}$.
\end{proof}

\begin{proof-of-lemma}[\ref{lem:preproc-prop}]
The claim immediately follows from applying
Lemma~\ref{lem:preproc-deterministic} with $\tclip = \frac{1}{\sqrt{8}}\bclip$
provided that the preconditions of the lemma hold with the desired probability.
That is, we need to show that with our desired probability, it holds that
\begin{enumerate}
  \item $\inf\{Z_i^T\scovinv{\uds}{}Z_i
    \mid i \in [n], \norm{Z_i}_\Sigma > \bclip\}
    > \frac{1}{4}\bclip^2$;
  \item $\dpsd(\scov{\bar{\uds}}{}, \Sigma) \leq \frac{1}{2}$;
  \item $\bar{\uds}$ is $(m + \eps n, \frac{1}{2})$-operator norm resilient,
    where $m = |\{i \in [n] \mid \norm{Z_i}_\Sigma \geq \bclip/8\}|$.
\end{enumerate}

By Lemma~\ref{lem:heavy-large-points}, the first event happens with
probability at least $1-C_km_kn^{1-k/2}$.

For the second event, we recall some useful properties we established in Section~\ref{sec:barP-props}.
In particular, because $\bclip \gtrsim m_k^{\frac{k}{k-2}}$,
we have that $\dpsd(\bar{\Sigma}, \Sigma) \leq \frac{1}{3}$ and
$\norm{\bar{Z}}_{\bar{\Sigma}}$ almost surely for $\bar{Z} \sim \bar{P}$.
This latter property implies by Lemma~\ref{lem:bdd-emp-cov-conc} that
$\dpsd(\scov{\bar{\uds}}{}, \bar{\Sigma})
= O\left(\sqrt{\frac{\bclip^2 \log(2d/\delta)}{n}}\right)$
with probability at least $1-\delta$.
Because we also assume that $n \gtrsim \bclip^2\log(d/\delta)$, we can combine
these errors to conclude that $\dpsd(\bar{\Sigma}, \Sigma) \leq \frac{1}{2}$
with probability at least $1-\delta$.

For the last event, we begin with a high probability upper bound on
$m = |\{i \in [n] \mid \norm{Z_i}_\Sigma \geq \bclip/8\}|$.
Note that $m$ is binomially-distributed with
$n$ draws and probability $\P(\norm{Z_i}_\Sigma \geq \bclip/8)$.
By Lemma~\ref{lem:heavy-norm-tail-bd}, we have that
\begin{align*}
  \P(\norm{Z_i}_\Sigma \geq \bclip/8)
  \leq \left(\frac{8m_k \sqrt{d}}{\bclip}\right)^k.
\end{align*}
Thus, a binomial tail bound yields
that with probability
at least $1-\exp\left(-\Omega\left( \left(\frac{8m_k \sqrt{d}}{\bclip}\right)^k n\right)\right)$, we have
$m \leq 2 \left(\frac{8 m_k \sqrt{d}}{\bclip}\right)^k n$.
We therefore need to show that with high probability,
$\bar{\uds}$ is
$(\mop, \frac{1}{2})$-operator norm resilient
for $\mop = 2 \left(\frac{8m_k \sqrt{d}}{\bclip}\right)^k n + \eps n$.
We have from Lemma~\ref{lem:heavy-op-norm-resilience} that
for $\mop \leq n$, there exists
\begin{align*}
  \aop
  = O\left(\sqrt{\frac{\bclip^2 \log(d/\delta)}{n}}
  + m_k^2\sqrt{\frac{d + \log(1/\delta)}{n}}
  + m_k^2 \left(\frac{\max\{\mop, d + \log(1/\delta)\}}{n}\right)^{1-\frac{2}{k}}\right)
\end{align*}
such that $\bar{\uds}$ is $(\mop, \aop)$-operator norm resilient with 
probability at least $1-\delta$.
Because $k\geq 4$ and $n \gtrsim \max\{\bclip^2\log(d/\delta), m_k^4(d+\log(1/\delta))\}$,
we can choose $\aop \leq \frac{1}{2}$ so long as
\begin{align*}
  m_k^2 \left(\frac{\mop}{n}\right)^{1-\frac{2}{k}}
  = m_k^2 \left(2 \left(\frac{8 m_k \sqrt{d}}{\bclip}\right)^k + \eps \right)^{1-\frac{2}{k}}
\end{align*}
is bounded above by a small constant.
This follows because
$\eps \lesssim m_k^{-\frac{2k}{k-2}}$
and $\bclip \gtrsim m_k^{\frac{k}{k-2}}\sqrt{d}$.

Combining all the failure probabilities, the claim follows.
\end{proof-of-lemma}
\subsection{Proof of Lemma~\ref{lem:heavy-goodness}}
\label{sec:pf-heavy-goodness}
We separately prove
operator norm resilience and inner product resilience in the next two lemmas,
and then put it all together afterwards to prove this lemma.

\begin{lemma}[Heavy-tailed operator norm resilience]
  \label{lem:heavy-op-norm-resilience}
  Suppose
  $n \gtrsim \bclip^2\log(2d/\delta)$ and $\mop \leq n$.
  There exists
  \begin{align*}
    \aop
    = O\left(\sqrt{\frac{\bclip^2 \log(d/\delta)}{n}}
    + m_k^2\sqrt{\frac{d + \log(1/\delta)}{n}}
    + m_k^2 \left(\frac{\max\{\mop, d + \log(1/\delta)\}}{n}\right)^{1-\frac{2}{k}}\right).
  \end{align*}
  such that
  with probability at least $1-4\delta$, $\bar{\uds}$ is
  $(\mop, \aop)$-operator norm resilient.
\end{lemma}
\begin{proof}
  Let $H \in \{\Hone, \Htwo\}$.
  For ease of notation, let $m = |H| = \frac{n}{2}$.
  We will prove the desired properties for $H$ hold with probability at least
  $1-2\delta$, so that the claim follows by a union bound over $H$.

  For a large constant $C > 0$,
  we may assume without loss of generality that
  $\mop \geq C(d + \log(1/\delta))$.
  This is because the claim for
  $\mop < C(d + \log(1/\delta))$ is weaker than the claim
  for $\mop = C(d + \log(1/\delta))$, possibly up to constant factors.

  We first show that $\scov{\bar{\uds}}{\genset} \preceq \frac{3}{4}\scov{\bar{\uds}}{}$.
  By Lemma~\ref{lem:bdd-emp-cov-conc} and \eqref{eq:Pbar-as-bd}, we have with
  probability at least $1-\delta$ that
  \begin{align*}
    \dpsd(\scov{\bar{\uds}}{}, \bar{\Sigma})
    = O\left(\sqrt{\frac{\bclip^2 \log(2d/\delta)}{n}}\right)
    ~~~\mbox{and}~~~
    \dpsd\left(\frac{n}{m}\scov{\bar{\uds}}{\genset}, \bar{\Sigma}\right)
    = O\left(\sqrt{\frac{\bclip^2 \log(2d/\delta)}{m}}\right).
  \end{align*}
  Because $n \geq m \gtrsim \bclip^2\log(2d/\delta)$, we have that
  $\max\{\dpsd(\scov{\bar{\uds}}{}, \bar{\Sigma}),
    \dpsd(\frac{n}{m}\scov{\bar{\uds}}{\genset}, \bar{\Sigma})\} \leq c$
  for a small constant $c > 0$.
  This rearranges to imply
  $\frac{n}{m}\scov{\bar{\uds}}{\genset} \preceq \frac{1+c}{1-c}\scov{\bar{\uds}}{}$.
  Because $m = \frac{n}{2}$, we can obtain the desired
  $\scov{\bar{\uds}}{S} \preceq \frac{3}{4}\scov{\bar{\uds}}{}$
  with probability at least $1-\delta$ by picking $c=\frac{1}{5}$.
  This establishes the first requirement for $(\mop, \aop)$-operator norm
  resilience.

  We now show the second requirement; that is, with probability at least $1-\delta$,
  we have for all $T \subseteq \genset$ with $|T| = \mop$ that
  $\scov{\bar{\uds}}{[n]\setminus F} \preceq \aop \scov{\bar{\uds}}{\genset}$.
  Because we are under the event that
  $\dpsd(\frac{1}{m}\sum_{i \in \genset} \bar{Z}_i \bar{Z}_i^T, \bar{\Sigma})
    \leq c$, we have that
  $\frac{1}{m}\sum_{i \in \genset} \bar{Z}_i \bar{Z}_i^T
    \succeq \frac{1}{2}\bar{\Sigma}$.
  Thus, it suffices to show that with probability at least $1-\delta$,
  it holds for all $T \subseteq \genset$
  with $|T| = \mop$ that
  $\frac{1}{m}\sum_{i \in T} \bar{Z}_i \bar{Z}_i^T \succeq \frac{\aop}{2}\bar{\Sigma}$.
  Equivalently, we show that for all such $T$ and all
  $v \in \S^{d-1} \cap \colspace(\bar{\Sigma})$,
  \begin{align*}
    \frac{1}{m}\sum_{i \in T} (v^T \bar{\Sigma}^{\dag/2} \bar{Z}_i)^2 \leq \frac{\aop}{2}.
  \end{align*}

  Let $\cV$ be a $\frac{1}{4}$-net of $\S^{d-1}$, which we can take such
  that
  $|\cV| \leq 9^d$ by \citet[Lemma 5.2]{Vershynin12}.
  Then \citet[Lemma 5.4]{Vershynin12} gives that
  \begin{align*}
    \sup_{v \in \S^{d-1}}
    \sup_{\substack{T \subseteq \genset \\|T|=\mop}}
    \frac{1}{m} \sum_{i \in T} (v^T \bar{\Sigma}^{\dag/2} \bar{Z}_i)^2
    \leq
    2\sup_{v \in \cV}
    \sup_{\substack{T \subseteq \genset \\|T|=\mop}}
    \frac{1}{m} \sum_{i \in T} (v^T \bar{\Sigma}^{\dag/2} \bar{Z}_i)^2.
  \end{align*}
  For each $v \in \cV$, we will bound
  \begin{align*}
    Y_v
    := \sup_{\substack{T \subseteq \genset \\|T|=\mop}}
    \frac{1}{m} \sum_{i \in T} (v^T \bar{\Sigma}^{\dag/2} \bar{Z}_i)^2
  \end{align*}
  with probability at least $1-9^{-d}\delta$,
  so that our desired claim will follow by a union bound over all $v \in \cV$.
  Observe that
  \begin{align*}
    Y_v
     & = \sup_{\substack{T \subseteq \genset                                                            \\ |T|=\mop}}
    \frac{1}{m}\sum_{i \in T} (v^T \bar{\Sigma}^{\dag/2} \bar{Z}_i)^2                               \\
     & = \frac{1}{m} \sum_{i \in \genset} (v^T \bar{\Sigma}^{\dag/2} \bar{Z}_i)^2
    - \inf_{\substack{T \subseteq \genset                                                               \\ |T|=m-\mop}}
    \frac{1}{m}\sum_{i \in T} (v^T \bar{\Sigma}^{\dag/2} \bar{Z}_i)^2                                                     \\
     & = \left[\frac{1}{m} \sum_{i \in \genset} (v^T \bar{\Sigma}^{\dag/2} \bar{Z}_i)^2 - 1 \right]
    - \frac{m-\mop}{m}\left[\inf_{\substack{T \subseteq S                                               \\ |T|=m-\mop}}
      \frac{1}{m-\mop}\sum_{i \in T} (v^T \bar{\Sigma}^{\dag/2} \bar{Z}_i)^2 - 1\right]
    + \frac{\mop}{m}.
  \end{align*}
  To bound the first term, we have
  \begin{align*}
    \left|\frac{1}{m} \sum_{i \in \genset}
    (v^T \bar{\Sigma}^{\dag/2} \bar{Z}_i)^2 - 1 \right|
     & = \left|v^T \bar{\Sigma}^{\dag/2} \left(\frac{n}{m}\scov{\bar{\uds}}{S}
    - \bar{\Sigma} \right) \bar{\Sigma}^{\dag/2} v \right|                        \\
     & \leq \dpsd\left(\frac{n}{m}\scov{\bar{\uds}}{\genset}, \bar{\Sigma}\right) \\
     & = O\left(\sqrt{\frac{\bclip^2 \log(2d/\delta)}{m}}\right).
  \end{align*}
  To bound the second term,
  we will use \citet[Proposition 3.14]{OliveiraRi22}, which we restate in
  Appendix~\ref{sec:heavy-helpers} as Lemma~\ref{lem:oliveira-prop}
  for convenience.
  This proposition gives
  subgaussian-like concentration for the truncated sample mean.
  We will apply Lemma~\ref{lem:oliveira-prop} with $\eta$ such that
  \begin{align*}
    \mop = \lfloor \eta m\rfloor + \lceil c\eta m + \log(4\cdot 9^d/\delta)\rceil.
  \end{align*}
  Because $\mop \geq C(d + \log(4/\delta))$ for a sufficiently large constant
  $C$, this $\eta$ will satisfy
  $\eta \asymp \frac{\mop}{m}$.
  Then by applying Lemma~\ref{lem:oliveira-prop} with this $\eta$,
  $q = \frac{k}{2}$,
  $\alpha = 9^{-d}\delta$, and
  $Y_i = W_i = (v^T \bar{\Sigma}^{\dag/2} \bar{Z}_i)^2$, we have with
  probability at least $1-\frac{9^{-d}}{2}\delta$ that
  \begin{align*}
     & \left| \inf_{\substack{T \subseteq \genset                                                                \\ |T|=m-\mop}}
    \frac{1}{m-\mop}\sum_{i \in T} (v^T \bar{\Sigma}^{\dag/2} \bar{Z}_i)^2 - 1 \right|                       \\
     & \qquad \lesssim \sqrt{\frac{\E[(v^T \bar{\Sigma}^{\dag/2} \bar{Z}_1)^4]\log(4 \cdot 9^d/\delta)}{m}}
    + \E[|v^T \bar{\Sigma}^{\dag/2} \bar{Z}_1|^{k}]^{\frac{2}{k}}\eta^{1 - \frac{2}{k}}                      \\
     & \qquad \lesssim m_k^2\sqrt{\frac{d + \log(1/\delta)}{m}}
    + m_k^2\left(\frac{\mop}{m}\right)^{1-\frac{2}{k}},
  \end{align*}
  where the last inequality is by the moment bound on $\bar{P}$ from
  \eqref{eq:Pbar-moment-bd}.
  Therefore we have,
  \begin{align*}
    Y_v
     & \lesssim
    \sqrt{\frac{\bclip^2 \log(2d/\delta)}{m}}
    + m_k^2\sqrt{\frac{d + \log(1/\delta)}{m}}
    + m_k^2 \left(\frac{\mop}{m}\right)^{1-\frac{2}{k}}
    + \frac{m}{n}                                         \\
     & \lesssim \sqrt{\frac{\bclip^2 \log(2d/\delta)}{m}}
    + m_k^2\sqrt{\frac{d + \log(1/\delta)}{m}}
    + m_k^2\left(\frac{\mop}{m}\right)^{1-\frac{2}{k}},
  \end{align*}
  proving the claim.
\end{proof}

\begin{lemma}[Heavy-tailed norm-boundedness]
  \label{lem:heavy-norm-bdd}
  Suppose
  $n \gtrsim \bclip^2\log(2d/\delta)$.
  With probability at least $1-2\delta$, $\bar{\uds}$ is
  $\frac{3}{\sqrt{2}}\bclip$-norm bounded.
\end{lemma}
\begin{proof}
  Let $H \in \{\Hone, \Htwo\}$.
  By Lemma~\ref{lem:bdd-emp-cov-conc} and \eqref{eq:Pbar-as-bd},
  we have with probability at least $1-\delta$,
  \begin{align*}
    \dpsd\left(\frac{2}{n}\sum_{i\in \genset} \bar{Z}_i \bar{Z}_i^T,
    \bar{\Sigma}\right)
    \leq \frac{1}{3}.
  \end{align*}
  Then by Lemma~\ref{lem:dpsd-prop}, we have that
  $\norm{\bar{\Sigma}^{1/2}
      (\frac{2}{n}\sum_{i\in \genset} \bar{Z}_i \bar{Z}_i^T)^{\dag}
      \bar{\Sigma}^{1/2}} \leq \frac{3}{2}$.
  Observe that
  \begin{align*}
    \scov{\bar{\uds}}{\genset}
    =\frac{1}{2} \cdot \frac{2}{n}\sum_{i\in \genset} \bar{Z}_i \bar{Z}_i^T,
  \end{align*}
  so that
  $\norm{\bar{\Sigma}^{1/2}\scovinv{\bar{\uds}}{\genset}\bar{\Sigma}^{1/2}} \leq 3$.

  Recall from \eqref{eq:Pbar-as-bd} that
  $\norm{\bar{Z}_i}_{\bar{\Sigma}} \leq \sqrt{\frac{3}{2}}\bclip$ almost surely, which implies
  \begin{align*}
    \bar{Z}_i^T\scovinv{\bar{\uds}}{S}\bar{Z}_i
    \leq \norm{\bar{\Sigma}^{1/2}\scovinv{\bar{\uds}}{\genset}\bar{\Sigma}^{1/2}}\norm{\bar{Z}_i}_{\bar{\Sigma}}^2
    \leq \frac{9}{2}\bclip^2,
  \end{align*}
  proving the claim after a union bound over $\genset$.
\end{proof}

\begin{lemma}\label{lem:inner-product-resilience}
  Suppose $n \gtrsim \bclip^2\log(2d/\delta_0)$ and
  that $\bar{\uds}$ is $(\mop,\aop)$-operator norm resilient
  with probability at least $1-\delta_0$. For $\gamma > 0$, let
  \begin{align*}
    \mip & \leq \mop \\
    \binner & \geq \sqrt{6} (1-\aop)^{-1}
    \bclip \cdot \gamma \\
    \delta  & \geq {|\genset|}^{\mip} (n-|\genset|)^{\mbig}
      \left(\sqrt{\frac{3}{2}} \cdot \frac{m_k}{\gamma}\right)^{k\mbig}.
  \end{align*}
  Then with probability at least $1-4\delta_0 - 2\delta$, it holds for all
  $\ipresset \subseteq [n]$ satisfying $|\ipresset| \leq \mip$ that
  $\bar{\uds}_{[n]\setminus \ipresset}$ is $(\mbig,\binner)$-inner product bounded.
\end{lemma}
\begin{proof}
We need to show that with probability at least $1-4\delta_0 - 2\delta$,
for all $\genset \in \{\Hone,\Htwo\}$
and for all $\ipresset \subseteq [n]$ satisfying $|\ipresset| \leq \mip$,
there exists $T \subseteq [n] \setminus \genset$ satisfying
$|T| \geq n - |\genset| - \mbig$
and
$|z_i^T \scovinv{\uds}{\genset \setminus \ipresset} z_j| \leq \binner$ for all distinct $i,j \in T \setminus \ipresset$.

We prove the claim with half the failure probability for $\genset = \Hone$,
so that a symmetric argument for $\genset = \Htwo$ and a union bound will prove the
desired claim.
Note that we may take $\ipresset \subseteq \genset$ without loss of generality,
as indices in $\ipresset \cap ([n] \setminus \genset)$ can be freely added to
$T$, which only loosens the requirements for $T$.

We begin with the following supporting lemma that shows it is unlikely
for too many disjoint pairs of indices to have a large inner product with
respect to the sample covariance.

\begin{lemma}\label{lem:inner-prod-resilience-helper}
  Let $\ipresset \subseteq \genset$ and let
  $\mc{A} \subset \{(i,j) \in ([n] \setminus \genset)^2 \mid i \neq j\}$
  such that every $i \in [n] \setminus \genset$ appears in at most one
  pair in $\mc{A}$.
  Let $A_{i,j}$ be the event that
  \begin{align}
    \absv{\bar{Z}_i^T \scovinv{\bar{\uds}}{[n]\setminus F} \bar{Z}_j}
    > \gamma \norm{\bar{\Sigma}^{1/2}\scovinv{\bar{\uds}}{[n]\setminus F} \bar{Z}_i}.
    \label{eqn:inner-prod-resilience-claim}
  \end{align}
  Then $\P(\cap_{(i,j) \in \mc{A}} A_{i,j})
    \leq \left(\sqrt{\frac{3}{2}} \cdot \frac{m_k}{\gamma}\right)^{k|\mc{A}|}$.
\end{lemma}
\begin{proof-of-lemma}[\ref{lem:inner-prod-resilience-helper}]
We first individually bound $\P(A_{i,j} \mid \bar{\uds}_{-j})$ for a each pair
$(i,j) \in \mc{A}$.
Then because $\scovinv{\bar{\uds}}{[n]\setminus F}\bar{Z}_i$ is independent of $\bar{Z}_j$ and also measurable
with respect to $\bar{\uds}_{-j}$, we have
\begin{align*}
  \P\left(|\bar{Z}_i^T \scovinv{\bar{\uds}}{[n]\setminus F} \bar{Z}_j|
  \geq \gamma \norm{\bar{\Sigma}^{1/2} \scovinv{\bar{\uds}}{[n]\setminus F} \bar{Z}_i} \mid \bar{\uds}_{-j}\right)
  \leq
  \sup_{v \in \S^{d-1}} \P(|\bar{\Sigma}^{\dag/2} \bar{Z}_j \cdot v| \geq \gamma).
\end{align*}
By the moment bound \eqref{eq:Pbar-moment-bd}, we have for every $v \in \S^{d-1}$ that
\begin{align*}
  \P(|\bar{\Sigma}^{\dag/2} \bar{Z}_j \cdot v| \geq \gamma)
  \leq \frac{1}{\gamma^k}\E[|\bar{\Sigma}^{\dag/2} \bar{Z}_j \cdot v|^k]
  \leq \left(\sqrt{\frac{3}{2}} \cdot \frac{m_k}{\gamma}\right)^k.
\end{align*}

From our assumptions that the pairs in $\mc{A}$ are disjoint and
$\mc{A} \subseteq ([n] \setminus \ipresset)^2$,
we obtain for all $(i,j) \in \mc{A}$ and
$(i',j') \in \mc{A} \setminus \{(i,j)\}$ that $A_{i',j'}$ is
measurable with respect to $\bar{\uds}_{-j}$.
The desired result then follows immediately.
\end{proof-of-lemma}

Now for all $\ipresset \subseteq \genset$ satisfying $|\ipresset| \leq \mip$
and all $\mc{A} \subseteq ([n] \setminus \genset)^2$ of $\frac{\mbig}{2}$
disjoint pairs of indices,
we apply Lemma~\ref{lem:inner-prod-resilience-helper}
with $T = \genset \setminus \ipresset$ and $\mc{A}$
and take a union bound.
As there are at most $|\genset|^{\mip}$ possible $\ipresset$
and at most $(n-|\genset|)^{\mbig}$ possible $\mc{A}$,
we have by our choice of $\delta$ that with probability at least $1-\delta$,
it holds for all $\ipresset$ and $\mc{A}$ that
there exists $(i, j) \in \mc{A}$ for which
\begin{align*}
  |\bar{Z}_i^T \scovinv{\bar{\uds}}{[n]\setminus F} \bar{Z}_j|
   & \leq
  \gamma\norm{\bar{\Sigma}^{1/2}\scovinv{\bar{\uds}}{[n]\setminus F} \bar{Z}_j}
  \leq \gamma\norm{\bar{\Sigma}^{1/2}\scovinv{\bar{\uds}}{[n]\setminus F}
    \bar{\Sigma}^{1/2}}
  \norm{\bar{\Sigma}^{\dag/2} \bar{Z}_j}.
\end{align*}
Now suppose that $\bar{\uds}$ is $(\mop, \aop)$-operator norm resilience with
respect to $\genset$,
which holds with probability at least $1-\delta_0$.
Then because we assume $\mip \leq \mop$, we have that
$\scov{\bar{\uds}}{[n]\setminus F} \succeq (1-\aop) \scov{\bar{\uds}}{\genset}$,
and so
$\norm{\bar{\Sigma}^{1/2}\scovinv{\bar{\uds}}{[n]\setminus F}\bar{\Sigma}^{1/2}}
  \leq (1-\aop)^{-1} \norm{\bar{\Sigma}^{1/2}\scovinv{\bar{\uds}}{\genset}\bar{\Sigma}^{1/2}}$.
By assumption on $n$ and Lemma~\ref{lem:bdd-emp-cov-conc}, we have that
$\norm{\bar{\Sigma}^{1/2}\scovinv{\bar{\uds}}{\genset}\bar{\Sigma}^{1/2}} \leq 2$
with probability at least $1-\delta_0$.
Furthermore, by \eqref{eq:Pbar-as-bd} we know that
$\norm{\bar{\Sigma}^{\dag/2} \bar{Z}_j} \leq \sqrt{\frac{3}{2}}\bclip$
by \eqref{eq:Pbar-as-bd} almost surely.

Therefore, by our choice of $\binner$,
we have that with probability at least $1-2\delta_0-\delta$, it holds for all
$\ipresset \subseteq \genset$ satisfying $|\ipresset| \leq \mip$ and
all $\mc{A} \subseteq ([n] \setminus \genset)^2$ of $\frac{\mbig}{2}$ disjoint pairs
of indices that there exists $(i, j) \in \mc{A}$ for which
$
  |\bar{Z}_i^T \scovinv{\bar{\uds}}{[n]\setminus F} \bar{Z}_j|
  \leq \binner
$.
Denote this event as $\mc{E}$.
We will now show that $\mc{E}$ will not hold if
there exists $\ipresset \subseteq \genset$ satisfying $|\ipresset| \leq \mip$
such that $\bar{\uds}_{[n]\setminus \ipresset}$ is not
$(\mbig, \binner)$-inner product bounded with respect to $\genset$,
proving the claim by contrapositive.

Suppose that $\bar{\uds}_{\ipresset}$ is not
$(\mbig, \binner)$-inner product bounded for some
$\ipresset \subseteq \genset$ satisfying $|\ipresset| \leq \mip$.
Then for all $T \subseteq [n] \setminus \genset$ with $|T| \geq n-|\genset|-\mbig$,
there exists distinct $i,j \in T$ for which
$|\bar{Z}_i^T\scovinv{\uds}{[n]\setminus \ipresset}\bar{Z}_j| > \binner$.
Now we iteratively construct a set $\mc{A}$.
First initialize $\mc{A}$ to $\emptyset$.
Then, until $\mc{A}$ consists of $\frac{\mbig}{2}$ pairs, iteratively find
two distinct
$i, j \in [n] \setminus \genset$
that don't appear in $\mc{A}$ and satisfies
$|\bar{Z}_i^T\scovinv{\uds}{[n]\setminus \ipresset}\bar{Z}_j| > \binner$
and add the pair $(i, j)$ to $\mc{A}$.
Note that finding such a pair is always possible, because while $\mc{A}$ has
less than $\frac{\mbig}{2}$ pairs, there are at least
$n-|\genset|-\mbig$ indices from which to choose $i$ and $j$,
and so our assumption implies there is a pair $(i, j)$ for which
$|\bar{Z}_i^T\scovinv{\uds}{[n]\setminus \ipresset}\bar{Z}_j| > \binner$.
This process results in $\mc{A}$ being a set of $\frac{\mbig}{2}$ disjoint pairs
of indices such that $|\bar{Z}_i^T\scovinv{\uds}{[n]\setminus \ipresset}\bar{Z}_j| > \binner$
for all $(i, j) \in \mc{A}$.
Thus, $\mc{E}$ does not hold, as desired.
\end{proof}

With these properties in hand, we now prove Lemma~\ref{lem:heavy-goodness}.

\begin{proof-of-lemma}[\ref{lem:heavy-goodness}]
Because $n \gtrsim \bclip^2\log(2d/\delta)$,
we have from Lemmas~\ref{lem:heavy-op-norm-resilience} that with
probability at least $1-O(\delta)$,
$\bar{\uds}$ is $(\mop, \aop)$-operator norm resilient
where
\begin{align*}
  \aop
  = O\left(\sqrt{\frac{\bclip^2 \log(d/\delta)}{n}}
  + m_k^2\sqrt{\frac{d + \log(1/\delta)}{n}}
  + m_k^2 \left(\frac{\max\{\mop, d + \log(1/\delta)\}}{n}\right)^{1-\frac{2}{k}}\right).
\end{align*}
Moreover, because $k \geq 4$ and
$n \gtrsim \max\{\bclip^2\log(d/\delta), m_k^4(d+\log(1/\delta))\}$,
this choice of $\aop$ satisfies $\aop \leq c$ for a small constant $c > 0$.
Increasing $n$ by a constant factor if necessary, we can also have with
probability at least $1-O(\delta)$
that $\scov{\bar{\uds}}{\Hone} \preceq \frac{3}{5}\scov{\bar{\uds}}{}$ and
$\scov{\bar{\uds}}{\Htwo} \preceq \frac{3}{5}\scov{\bar{\uds}}{}$, rather than just
$\frac{3}{4}\scov{\bar{\uds}}{}$.
We also have by Lemma~\ref{lem:heavy-norm-bdd} that
$\bar{\uds}$ is $O(\bclip)$-norm bounded with probability at least $1-O(\delta)$.

Under these events, we first show that for all $\ipresset \subseteq [n]$ such that
$|\ipresset| \leq \mip$,
the dataset $\bar{\uds}_{[n] \setminus \ipresset}$ is
$O(\bclip)$-norm bounded and
$(\mop, O(\aop))$-operator norm resilient.
We will only show the required properties hold with respect to $H = \Hone$, as the
case where $H = \Htwo$ is symmetric.

For $O(\bclip)$-norm boundedness, we have for all $i \in [n]$ that
\begin{align*}
  \bar{Z}_i^T\scovinv{\bar{\uds}}{\Hone \setminus \ipresset}\bar{Z}_i
  \leq (1-\aop)^{-1}\bar{Z}_i^T\scovinv{\bar{\uds}}{\Hone}\bar{Z}_i
  = O(\bclip^2),
\end{align*}
where the first inequality is because $|\ipresset| \leq \mip \leq \mop$ and
$(\mop, \aop)$-inner product resilience
and the final bound is by $O(\bclip)$-norm boundedness.

For $(\mop, O(\aop))$-operator norm resilience,
we first show that
\begin{align*}
  \scov{\bar{\uds}}{\Hone\setminus \ipresset}
  \preceq \scov{\bar{\uds}}{\Hone}
  \preceq \frac{3}{5}\scov{\bar{\uds}}{}
  \preceq \frac{3}{5(1-\aop)}\scov{\bar{\uds}}{[n] \setminus \ipresset},
\end{align*}
where the last inequality is by operator norm resilience.
Because $\aop \leq c$ for a sufficiently small constant $c$,
we have that
$\frac{3}{5(1-\aop)} \leq \frac{3}{4}$ and so
$\scov{\bar{\uds}}{\Hone\setminus \ipresset} \preceq \frac{3}{4}\scov{\bar{\uds}}{[n]\setminus \ipresset}$.
Now for any $T \subseteq \Hone$ such that $|T| \leq \mop$,
\begin{align*}
  \scov{\bar{\uds}}{T \setminus \ipresset}
  \preceq \scov{\bar{\uds}}{T}
  \preceq \aop \scov{\bar{\uds}}{\Hone}
  \preceq \frac{\aop}{1-\aop}\scov{\bar{\uds}}{\Hone\setminus \ipresset}
  \preceq O(\aop)\scov{\bar{\uds}}{\Hone\setminus \ipresset}.
\end{align*}

Finally we establish the required inner product boundedness properties.
By applying Lemma~\ref{lem:inner-product-resilience}
with $\gamma = \sqrt{\frac{3}{2}}m_k \cdot \left(\frac{n}{2}\right)^{1/k}$,
we have that $\bar{\uds}_{[n]\setminus \ipresset}$ is
$(\mbig, O(m_k n^{1/k} \bclip))$-inner product bounded for all
$\ipresset \subseteq [n]$ satisfying $|\ipresset| \leq \mip$, simultaneously
with probability at least
\begin{align*}
  1 - O\left(\left(\frac{n}{2}\right)^{\mip + \mbig - k\mbig}\right)
  = 1 - O\left(n^{-\Omega(k\mbig)}\right).
\end{align*}
The last equality above is because we assume $\mip \leq \mbig$ and $k \geq 4$.

This proves the claim.
\end{proof-of-lemma}
\subsection{Proof of Corollary~\ref{cor:heavy-alg-perf}}\label{sec:heavy-alg-perf-pf}
Applying Lemma~\ref{lem:preproc-prop}, we have
with probability at least $1-C_k m_k^k n^{1-k/2} - 2\delta - \exp\left(-\Omega\left( \left(\frac{8m_k\sqrt{d}}{\bclip}\right)^k n\right)\right)$,
there exists $T \subseteq [n]$ with
$|T| \leq 2 \left(\frac{8 m_k \sqrt{d}}{\bclip}\right)^k n$
such that $\cY$ is a
$\eps'$-contamination of
$\bar{\uds}_{[n]\setminus T}$.

Then because
$\left(\frac{8 m_k \sqrt{d}}{\bclip}\right)^k \lesssim \gamma \lesssim m_k^{-\frac{2k}{k-2}}$,
we can apply Lemma~\ref{lem:heavy-goodness}
with
$\mbig \geq 2 \left(\frac{4 m_k \sqrt{d}}{\bclip}\right)^k n$ and $\mop =
\gamma n$
to get that there exists
\begin{align*}
  \bnorm  & = O(\bclip)                                      \\
  \binner & = O(m_k n^{1/k}\bclip)                           \\
  \aop    & = O\left(\sqrt{\frac{\bclip^2\log(d/\delta)}{n}}
  + m_k^2 \gamma^{1-\frac{2}{k}}\right).
\end{align*}
such that
both $\bar{\uds}$ and 
$\bar{\uds}_{[n]\setminus T}$
are $(\bnorm, \mbig, \binner, \mop, \aop)$-good
with probability at least $1-O(\delta + n^{-\Omega(k\mbig)})$.

For ease of notation, let $\eps' = \frac{9n}{\bclip^2}\eps$.
To apply Theorem~\ref{thm:alg-result} to
$\SigmaAlg = \alg(\cY, \eps', \tinner, \tnorm)$, we need to choose the
parameters $\tinner, \tnorm$ so that
\begin{enumerate}
  \item $\eps' \leq \frac{1}{4(1+C)}$,
  \item $\tnorm \geq (1-\aop)^{-\frac{1}{2}}\bnorm$,
  \item $\tinner \geq \max\{2\binner, \frac{8e\aop\log(\tnorm)}{1-\aop} \cdot \frac{1}{\eps'}\}$,
  \item $n \geq \frac{12e\tnorm^2\log(\tnorm)}{\eps'\tinner}$,
  \item $\mop \geq \max\{1 + C, 1 + \frac{72(1+C)\tnorm^2}{\tinner}\}\eps' n + 4\mbig$.
\end{enumerate}
Because $\gamma \gtrsim \left(\frac{8 m_k \sqrt{d}}{\bclip}\right)^k$, we have that $\mop \geq 8\mbig$, so the fifth requirement above amounts to
$\eps' \lesssim \gamma$ (which is satisfied by assumption on $\gamma$)
and $\tinner \gtrsim \frac{\tnorm^2\eps'}{\gamma}$.

Then picking $\tnorm = (1-\aop)^{-\frac{1}{2}}\bnorm$, the only remaining
parameter to choose is $\tinner$ and we need it to satisfy
\begin{align}
  \tinner
  \gtrsim \max\{\binner,
  \frac{\aop\log(\tnorm)}{(1-\aop)\eps'},
  \frac{\tnorm^2\log(\tnorm)}{\eps' n},
  \frac{\tnorm^2 \eps'}{\gamma}
  \}.\label{eq:heavy-tinner-lb-1}
\end{align}

Because $n \gtrsim \bclip^2\log(d/\delta)$ and $\mop \lesssim m_k^{-\frac{2k}{k-2}}n$, we have that $\aop \leq \frac{1}{2}$.
This means satisfies $\tnorm \asymp \bclip$ and we can
simplify \eqref{eq:heavy-tinner-lb-1} to get
\begin{align*}
  \tinner
  \gtrsim \max\{
  m_k n^{1/k} \bclip,
  \frac{\aop \log \bclip}{\eps'},
  \frac{\bclip^2\log \bclip}{\eps' n},
  \frac{\eps' \bclip^2}{\gamma}
  \}.
\end{align*}
With these parameters, we may apply Theorem~\ref{thm:alg-result}
to obtain
\begin{align*}
  \err(\SigmaAlg, \scov{\bar{\uds}}{[n]\setminus T})
  \lesssim \aop + \eps' \tinner.
\end{align*}
Moreover, by operator norm resilience of $\bar{\uds}$,
\begin{align*}
  \err(\scov{\bar{\uds}}{[n]\setminus T}, \scov{\bar{\uds}}{}) = \aop.
\end{align*}
Putting everything together, we have
\begin{align*}
  \err(\SigmaAlg, \scov{\bar{\uds}}{})
   & \lesssim \aop + \eps' \tinner    \\
   & \lesssim m_k\eps' n^{1/k} \bclip
  + \log \bclip \left(\sqrt{\frac{\bclip^2 \log(d/\delta)}{n}}
  + m_k^2 \gamma^{1-\frac{2}{k}}\right) \\
  &\qquad+ \frac{\bclip^2\log \bclip}{n}
  + \frac{(\eps')^2 \bclip^2}{\gamma}.
\end{align*}

Recalling that $\eps' = \frac{9\eps n}{\bclip^2}$ and ignoring
lower-order terms, we finally have
\begin{align*}
  \err(\SigmaAlg, \scov{\bar{\uds}}{})
  &\lesssim \log \bclip\sqrt{\frac{\bclip^2 \log(d/\delta)}{n}}
   + \frac{m_k \eps n^{1 + \frac{1}{k}}}{\bclip}
   + m_k^2 \gamma^{1-\frac{2}{k}}\log \bclip
   + \frac{\eps^2 n^2}{\bclip^2 \gamma},
\end{align*}
as desired.


\end{document}